\documentclass[jkps,preprint,fleqn,showpacs,showkeys]{revtex4}
\usepackage{graphicx}
\usepackage{amssymb}
\usepackage{amsmath}
\usepackage{bm}

\newcommand{\etal}{{\it et al.~{}}}

\newcommand{\refeq}[1]{Eq.~(\ref{eq:#1})}
\newcommand{\fig}[1]{\label{fig:#1}}
\newcommand{\reffig}[1]{Fig.~\ref{fig:#1}}

\newcommand{\lasec}[1]{\label{sec:#1}}
\newcommand{\refsec}[1]{Sec.~\ref{sec:#1}}

\newcommand{\beq}{\begin{eqnarray}}   
\newcommand{\eeq}{\end{eqnarray}}
\newcommand{\nn}{\nonumber}

\def \d {{\partial}}

\newcommand{\ket}[1]{| #1 \rangle} 

\begin{document}
\setcounter{page}{0}


\title{Aspects of Superfluid Cold Atomic Gases in Optical Lattices
}

\author{Gentaro Watanabe}
\affiliation{Asia Pacific Center for Theoretical Physics (APCTP),
Pohang, Gyeongbuk 790-784, Korea,}
\affiliation{Department of Physics, Pohang University of Science and Technology (POSTECH),
Pohang, Gyeongbuk 790-784, Korea,}
\affiliation{Nishina Center, RIKEN, 2-1 Hirosawa, Wako, Saitama 351-0198, Japan}

\author{Sukjin Yoon}
\affiliation{Asia Pacific Center for Theoretical Physics (APCTP),
Pohang, Gyeongbuk 790-784, Korea}

\date{\today}

\begin{abstract}
We review our studies on Bose and Fermi superfluids of cold atomic
gases in optical lattices at zero temperature.  Especially, we focus
on superfluid Fermi gases along the crossover between the
Bardeen-Cooper-Schrieffer (BCS) and the Bose-Einstein condensate (BEC)
states, which enable us to study the Bose and the Fermi superfluids in a
unified point of view.  We discuss basic static and long-wavelength properties (such as the equation of state, incompressibility, and effective mass), energetic
stability, and energy band structures of the superfluid Fermi gases in
an optical lattice periodic along one spatial direction.  The periodic
potential causes pairs of atoms to be strongly bound, and this can
affect the static and long-wavelength properties and the stability of the
superflow.  Regarding the band structure, a peculiar loop structure
called ``swallowtail'' can appear in superfluid Fermi gases and 
in the Bose case, but the mechanism of emergence in the Fermi case is very
different from that in bosonic case.  
Other quantum phases that the cold
atomic gases in optical lattices can show are also briefly discussed
based on their roles as quantum simulators of Hubbard models.
\end{abstract}


\maketitle

\section{INTRODUCTION}

In 1995, the atomic physics community witnessed the coolest state of matter
realized in the laboratory by that time.  With lasers and
electromagnetic fields, atomic Bose gases were trapped and cooled 
to within a micro-Kelvin to make them coalesce into a single quantum
wave, known as a Bose-Einstein condensate (BEC) state \cite{bec_exp}.
With the development and upgrading of technology and tools, finally about 10
years later, physicists succeeded in cooling atomic Fermi gases 
to even lower temperatures and in providing conclusive evidence
that the interacting atomic Fermi gas could be controlled to be in a
state from a Bardeen-Cooper-Schrieffer (BCS) superfluid of loosely-bound
pairs to a BEC of tightly-bound dimers \cite{fermi_superfluid}.  This
quantum wonderland of cold atomic gases has been proven to be rich in
physics, and it has also shown us new realms of
research\ \cite{rev_bec_leggett,rev_trapped_bec,rev_bloch,giorgini,quantum simulator}.  
Unprecedented high-precision control over the cold
atomic gases has made it possible to mimic various quantum systems.
It has also paved a way to new physical parameter region that had
not been attainable in other quantum systems and, therefore, had not
been well thought about.

In a cold Fermi gas with an equal mixture of two hyperfine states 
(for simplicity, called spin up/down), 
wide-range control of the short-range interatomic interaction
(characterized by an $s$-wave scattering length $a_s$) via
Feshbach resonances \cite{fano_feshbach} has enabled the mapping of a new landscape of superfluidity, 
known as ``BCS-BEC crossover,'' 
smoothly connecting the BCS superfluid to the BEC superfluid \cite{bcs_bec_crossover}.
For a weakly-attractive interaction ($1/k_{\rm F} a_s \ll -1$, 
with $k_{\rm F}$ conventionally being the Fermi momentum of a free Fermi gas of the same density), the gas shows a
superfluid behavior originating from the many-body physics of Cooper
pairs that are formed from weak pairings of atoms of opposite spin and
momentum and whose spatial extent is larger than the interparticle
distance. For a strongly-attractive interaction ($1/k_{\rm F} a_s \gg 1$), the
gas shows a superfluid behavior that can be explained by the
Bose-Einstein condensation of tightly-bound bosonic 
dimers made up of two fermions of opposite spin. In the crossover
region/unitary region ($1/k_{\rm F} |a_s| \alt 1$) bridging the two
well-understood regions above, the gas is strongly interacting, which
means that the interaction energy is comparable to the Fermi energy of
a free Fermi gas of the same density and defies perturbative
many-body techniques. Especially, the physics in the so-called
``unitary limit'' ($|a_s|=\infty$) is universal because only one
remaining length scale, $1/k_{\rm F}$, which is approximately the
interparticle distance, appears in the equation of state of the gas,
while the interaction effect appears as a universal dimensionless
parameter. 
Because the physics at unitarity is universal, i.e., 
independent of its constituents and is non-perturbative due to the strong interaction, 
the unitary Fermi gas has been the test-bed for the various theoretical techniques 
developed so far (see, e.g., Sec.~V\,B and V\,C in Ref.~\onlinecite{giorgini}).

Cold atomic gases also provide insights into other forms of matter.
Inside neutron stars, there might be quark matter made of different
ratios of quarks, and this imbalance might result in new types of
superfluid, such as the Fulde-Ferrell-Larkin-Ovchinnikov (FFLO) phase,
which has spatially non-uniform order parameter \cite{FFLO}.  Even
though the electric-charge neutrality of atoms might change the
picture, we might get some ideas from investigations on imbalanced
atomic Fermi gases \cite{imbalance}.  Another example is the
quark-gluon plasma (QGP) created in the Relativistic Heavy Ion
Collider (RHIC).  A QGP with a temperature of about $10^{12}$ K was produced by
smashing nuclei together.  Measurements on its expansion after its
creation show that the QGP is a nearly perfect fluid with very small shear
viscosity.  A cloud of atomic Fermi gases at unitarity shows the same
strange behavior while the origins of the similarity between hot
and cold matter are to be speculated \cite{hot_cold}.

Cold atoms in periodic potentials are more intriguing because of their possible connection to the solid state/condensed matter physics
(see, e.g., Refs.\ \onlinecite{morsch,rev_bloch}, and \onlinecite{yukalov} for reviews). The artificial periodic potential of light, so-called ``optical lattice,'' is configured with pairs of counter-propagating laser fields that are detuned from atomic transition frequencies enough that they act as practically free-of-defect, conservative potentials that the atoms experience via the optical dipole force. 
The geometry and the depth of the lattice potential can be controlled 
by orienting the directions and by changing the intensities of the laser beams, respectively.
Moreover, the
macroscopically-large lattice constant of the optical lattice compared
with the lattice potential in solids is a great advantage for
experimental observation and manipulation; nowadays, {\it in-situ}
imaging and addressing of strongly-correlated systems at the
single-site level have been realized (e.g.,
Ref.\ \onlinecite{single_site}).
The high controllability of both the optical lattice properties via laser fields and the interatomic interaction via the Feshbach resonance allows cold atomic gases to serve as quantum simulators for various models of condensed matter physics. 
Many phenomena of solid state/condensed matter physics have already been observed or realized in cold atomic gases on the optical lattices, including a band structure, a Bloch oscillation, Landau-Zener tunneling,
and a superfluid-Mott insulator transition
\cite{bloch_zener,Greiner_SF_Mott}.
Precision control and measurement at the level of one lattice site enable the cold atomic gases on optical lattices to be applied in inventing new manipulation techniques and novel devices, such as matter-wave interferometers, optical lattice clocks, and quantum registers \cite{optlatclock,jaksch}.

In this review article, we mainly explore the superfluid properties of
cold atomic gases in an optical lattice at zero temperature.  
Superfluidity is the most
well-known quantum phase of ultracold atomic gases and is prevalent in
many other systems, such as superconductors, superfluid helium, and
superfluid neutrons in ``pasta'' phases (see, e.g., Ref.\ \onlinecite{pasta})
of neutron star crusts. Knowing its properties is also important both for
judging whether the superfluid state is approached in experiments and
applying its properties in controlling the system.  Depending on the
physical conditions, cold atomic gases in optical lattices show
various interesting phases other than the superfluid phase, for
example, simulating Hubbard models and some other quantum phases will be explained
very briefly in \refsec{others}.
In our discussion, we mainly consider atomic Fermi gases 
partly because weakly-interacting atomic Bose gases can be considered 
as an extreme limit in the BCS-BEC crossover of the atomic Fermi gases
and partly because the physics of the cold Fermi gases is richer. 
Bose gases are discussed separately in the cases where 
we cannot find the corresponding phenomena in the Fermi gases. 
This does not mean that the cold Bose gases are less interesting and less important.
The tools and the techniques developed for creating the BEC were the building blocks 
for those for creating superfluid Fermi gases, 
and superfluid properties of the Bose gases might be more useful in some applications 
due to their higher superfluid transition temperature.

In \refsec{formalism}, our system and the theoretical frameworks are
presented.  There, we explain the mean-field theory and the
hydrodynamic theory, as well as the validity region of each theory.
Using cold atomic gases, various parameter regions become accessible,
and the mean-field theory in the continuum model is a powerful tool
that allows us to study superfluid states covering such a wide
region. Hydrodynamic theory, though its validity region is limited,
can provide precise analytic predictions in some limits, which are
complementary to mean-field theory.  Based on these
methods, we study various aspects of the superfluid state in a lattice
potential periodic along one spatial direction. This setup is the
simplest, but contains essential effects of the lattice potential.

Section \ref{sec:thermo} discusses basic static and long-wavelength
properties of the cold Fermi gases in an optical
lattice. Specifically, the incompressibility and the effective mass
are obtained from the equation of state (EOS). Focusing on the unitary
Fermi gas as a typical example, here we show that these properties in
the lattice can be strongly modified from those in free space
\cite{optlatunit}. Formation of bound pairs is assisted by the
periodic potential, and this results in a qualitative change of the
EOS.

In \refsec{stability}, we examine the stability of the superfluid flow
in the optical lattice. Mainly, the energetic instability and the
corresponding critical velocity of the superflow are discussed.
The stability of the superflow is a fundamental problem of the
superfluidity, and cold atomic gases allow us to study this important
problem for various parameter regions for both the Bose and the Fermi
superfluids.  Unlike an obstacle in uniform superfluids, the periodic
potential can modify the excitation spectrum, and this manifests itself
in the behavior of the critical velocity \cite{vc,vc_crossover}.

In \refsec{energy_band}, we investigate the energy band structures of
cold atomic Fermi gases in an optical lattice along the BCS-BEC
crossover.  By tuning the interatomic interaction strength, we can make the
nonlinear effect of the interaction energy to dominate
the external periodic potential.  In this parameter region,
``swallowtail'' energy band structures emerge, and the physical
properties of the system are affected by their appearance
\cite{swallowtail}.  This is one example of cold atomic gases 
showing physical phenomena inaccessible in other systems.

In \refsec{others}, we discuss quantum phases of cold atomic gases in
optical lattices and show how they play their role as ``quantum
simulators'' of Hubbard models.  Before the advent of cold atomic
gases in optical lattices, the Hubbard model was considered as an
approximate toy model of more complicated real systems and was one of
the central research topics in solid-state/condensed-matter physics.
High-precision control of cold atomic gases in optical lattices
now allows Hubbard models to be realized in experiments, so 
open questions of quantum magnetism and high $T_c$
superconductivity can be addressed.
Finally, \refsec{summary} contains a summary and presents some perspectives on the cold atomic gases in optical lattices.

\section{THEORETICAL FRAMEWORKS\lasec{formalism}}

\subsection*{2.1.\quad Setup of the System and the Periodic Potential\label{sec:setup}}

In the present article, we discuss
superfluid flow made of either fermionic or bosonic cold atomic gases
subject to an optical lattice.
For concreteness, we consider one of the most typical cases:
the system is three dimensional (3D), and the 
potential is periodic in one dimension 
with the following form:
\begin{equation}
  V_{\rm ext}({\bf r})=sE_{\rm R} \sin^2{q_B z} \equiv V_0 \sin^2{q_B z} .
\end{equation}
Here, $V_0\equiv sE_{\rm R}$ is the lattice height, 
$s$ is the lattice intensity in dimensionless units, 
$E_{\rm R}=\hbar^2q_{\rm B}^2/2m$ is the recoil energy, 
$m$ is the mass of atoms,
$q_{\rm B}=\pi/d$ is the Bragg wave vector, and $d$ is the lattice constant.
For simplicity, we also assume that the supercurrent is 
in the $z$ direction; thus, the system is uniform in the transverse
(i.e., $x$ and $y$) directions.
Throughout the present article, we set the temperature $T=0$.

Before giving a detailed description of the theoretical framework,
it is useful to summarize the scales in this system.
The periodic potential is characterized by two energy scales:
one is the recoil energy $E_{\rm R}$, which is directly related to 
the lattice constant $d$, and the other is the lattice height $V_0=sE_{\rm R}$.
Regarding Bose-Einstein condensates of bosonic atoms, 
a characteristic energy scale is the interaction energy $gn$, where
$g= 4\pi\hbar^2a_s/m$ is the interaction strength
and $n$ is the density of atoms.
On the other hand, in the case of superfluid Fermi gases,
the total energy is on the order of the Fermi energy
$E_{\rm F}= \hbar^2k_{\rm F}^2/(2m)$, 
with the Fermi wave number $k_{\rm F}\equiv (3\pi^2 n)^{1/3}$
corresponding to that of a uniform non-interacting
Fermi gas with the same density $n$.
Therefore, the relative effect of the lattice strength
is given by the ratio $\eta_{\rm height}=V_0/gn$ for bosons and 
$\eta_{\rm height}=V_0/E_{\rm F}$
for fermions.  
Likewise, the relative fineness of the lattice 
(compared to the healing length)
is characterized by 
the ratio $\eta_{\rm fine}=(gn/E_{\rm R})^{-1}\sim (\xi/d)^2$ for bosons and 
$\eta_{\rm fine}=(E_{\rm F}/E_{\rm R})^{-1} \sim (k_{\rm F} d)^{-2}$ for fermions, 
which is $\sim (\xi/d)^2$ near unitarity.
Here, $\xi$ is the healing length given as
$\xi=\hbar/(2mgn)^{1/2}$ for Bose superfluids and as $\xi\sim k_{\rm F}^{-1}$
for Fermi superfluids at unitarity, which is consistent with the BCS
coherence length $\xi_{\rm BCS}=\hbar v_{\rm F}/\Delta$, where
$v_{\rm F}=\hbar k_{\rm F}/m$ and $\Delta$ is the pairing gap.
We can also say that the validity of the local density approximation (LDA)
is characterized by $1/\eta_{\rm fine}\agt (d/\xi)^2 \gg 1$
corresponding to a lattice with a low fineness $\eta_{\rm fine}\ll 1$.
In the present article, we shall consider a large parameter region
covering weak to strong lattices, 
$\eta_{\rm height}=O(10^{-2})$ -- $O(10)$\
($s\sim 0.1$ -- $5$),
and low to high fine lattices,
$\eta_{\rm fine}\sim 0.1$ -- $10$.

\subsection*{2.2.\quad Mean-field Theory in the Continuum Model}

For the study the superfluidity in such various regions 
in a unified manner, one of the most useful theoretical frameworks is the
mean-field theory in the continuum model.
Because our system is a superfluid in three dimensions and 
the number of particles in each site is infinite in our setup,
the effects of quantum fluctuations, which are not captured by the 
mean-field theory, may be small.
We also note that widely-used tight-binding models are 
invalid in the weak lattice region of $\eta_{\rm height}\alt 1$.

For dilute BECs at zero temperature, the system is well described by
the Gross-Pitaevskii (GP) equation \cite{gross,pitaevskii61,rev_trapped_bec}:
\begin{equation}
- \frac{\hbar^2}{2m}\partial_{z}^2 \Psi + V_{\rm ext}(z) \Psi +
g | \Psi |^2 \Psi = \mu \Psi ,
\label{eq:gp}
\end{equation}
where $\Psi(z)=\sqrt{n(z)}\exp[i\phi(z)]$ is the 
condensate wave function, $\phi(z)$ is its phase, and $\mu$ is
the chemical potential.
The local superfluid velocity is given by
$v(z)=(\hbar/m)\partial_z \phi(z)$.

For superfluid Fermi gases, we consider a balanced system of
attractively-interacting (pseudo)spin $1/2$ fermions, 
where the density of each spin component is $n/2$.
To describe the BCS-BEC crossover at zero temperature,
we use the Bogoliubov-de Gennes (BdG) equations 
\cite{bdg,giorgini}
\begin{equation}
\left( \begin{array}{cc}
H'(\mathbf r) & \Delta (\mathbf r) \\
\Delta^\ast(\mathbf r) & -H'(\mathbf r) \end{array} \right)
\left( \begin{array}{c} u_i( \mathbf r) \\ v_i(\mathbf r)
\end{array} \right)
=\epsilon_i\left( \begin{array}{c} u_i(\mathbf r) \\
v_i(\mathbf r) \end{array} \right) \; ,
\label{eq:BdG}
\end{equation}
with $H'(\mathbf r) =-\hbar^2 \nabla^2/2m +V_{\rm ext}-\mu$,
$u_i(\mathbf r)$ and $v_i(\mathbf r)$ are quasiparticle wave functions,
which obey the normalization condition 
$\int d{\mathbf r}\ [u_i^*(\mathbf r)u_j(\mathbf r)+v_i^*(\mathbf r)v_j(\mathbf r)]=\delta_{i,j}$,
and $\epsilon_i$ are the corresponding quasiparticle energies. 
The pairing field $\Delta(\mathbf r)$ and the chemical potential $\mu$ 
in Eq.\ (\ref{eq:BdG})
are self-consistently determined from the gap equation
\begin{equation}
  \Delta(\mathbf r) =-g \sum_i u_i(\mathbf r) v_i^*(\mathbf r) \; ,
\label{eq:gap}
\end{equation}
together with the constraint on the average number density
\begin{equation}
  \bar{n}=\frac{2}{V} \sum_i \int |v_i(\mathbf r)|^2\ d{\mathbf r}
=\frac{1}{V}\int n({\bf r})\ d{\mathbf r},
\end{equation}
with $n({\bf r})\equiv 2 \sum_i |v_i({\bf r})|^2$.
Here, $g$ is the coupling constant for the contact interaction, and 
$V$ is the volume of the system. 
The average energy density $\bar{e}$ can be calculated as
\begin{equation}
\bar{e} = \frac{1}{V}\int \! d{\bf r} \Biggl[\frac{\hbar^2}{2m}
\left(2\sum_i|\bm{\nabla}v_i|^2\right) + V_{\rm ext}\, n({\bf r})
+\frac{1}{g} |\Delta({\bf r})|^2 \Biggr]\, .
\label{eq:energydens}
\end{equation}

For contact interactions, the right-hand side of Eq.\ (\ref{eq:gap})
has an ultraviolet divergence, which has to be regularized by
replacing the bare coupling constant $g$ with the two-body $T$-matrix
related to the $s$-wave scattering length
\cite{randeria,bruun,bulgac,grasso}.
A standard scheme \cite{randeria} is to introduce the cutoff energy 
$E_C\equiv\hbar^2k_C^2/2m$ 
in the sum over the BdG eigenstates and to replace $g$ by the following relation:
\begin{equation}
\frac{1}{g}=\frac{m}{4\pi\hbar^2 a_s}-\sum_{k<k_C} 
\frac{1}{2\epsilon^{(0)}_k},
\end{equation}
with $\epsilon^{(0)}_k\equiv \hbar^2k^2/2m$.

In the presence of a 
supercurrent with wavevector $Q=P/\hbar$ 
($P$ is the quasi-momentum for atoms rather than for pairs; thus, it is defined
in the range of $|P| \le \hbar q_{\rm B}/2$)
in the $z$ direction,
one can write the quasiparticle wavefunctions in the Bloch form as
$u_i(\mathbf r) =
\tilde{u}_i(z) e^{i Q z}e^{i\mathbf k \cdot \mathbf r }$ and
$v_i(\mathbf r) = \tilde{v}_i(z) e^{-i Q z}e^{i\mathbf k \cdot
\mathbf r }$,
leading to the pairing field 
\begin{equation}
  \Delta(\mathbf r)=e^{i 2Q z}\tilde{\Delta}(z) .
\end{equation}
Here, $\tilde{\Delta}(z)$, 
$\tilde{u}_i(z)$, and $\tilde{v}_i(z)$ 
are complex functions with period $d$, and
the wave vector $k_z$ ($|k_z| \le q_{\rm B}$) lies in the 
first Brillouin zone. This Bloch decomposition transforms 
Eq.~(\ref{eq:BdG}) into the following BdG equations for $\tilde{u}_i(z)$ 
and $\tilde{v}_i(z)$ :
\begin{equation}
\left( \begin{array}{cc}
\tilde{H}_{Q}(z) & \tilde{\Delta}(z) \\
\tilde{\Delta}^\ast(z) & -\tilde{H}_{-Q}(z) \end{array} \right)
\left( \begin{array}{c} \tilde{u}_i(z) \\ \tilde{v}_i(z)
\end{array} \right)
=\epsilon_i\left( \begin{array}{c} \tilde{u}_i(z) \\
\tilde{v}_i(z) \end{array} \right) \;,
\label{eq:BdG2}
\end{equation}
where
\begin{equation}
  \tilde{H}_{Q}(z)\equiv \frac{\hbar^2}{2m} \left[ k^2_\perp
+\left(-i\partial_z+Q+k_z\right)^2 \right] +V_{\rm ext}(z) -\mu\, .
\nonumber\label{hq}
\end{equation}
Here, $k_\perp^2\equiv k_x^2 + k_y^2$, and
the label $i$ represents the wave vector $\mathbf k$, as well 
as the band index.

\subsection*{2.3.\quad Hydrodynamic Theory\label{sec:hydro}}

When the local density approximation (LDA) is valid, such that 
the typical length scale of the density variation given by $d$
is much larger than the healing length $\xi$ of the superfluid,
hydrodynamic theory in the LDA can be useful \cite{vc}.
In hydrodynamic theory, we describe the system 
in terms of the density field $n(z)$ and the (quasi-)momentum field $P(z)$
[or the velocity field $v(z)$].
The LDA assumes that, locally, the system behaves like a uniform
gas; thus, the energy density $e(n,P)$ can be written in the form
\begin{equation}
  e(n,P)=nP^2/2m + e(n,0),
\end{equation}
and one can define the local chemical
potential $\mu(n)=\partial e(n,0) /\partial n$. The density
profile of the gas at rest in the presence of the external
potential can be obtained from the Thomas-Fermi relation $\mu_0 =
\mu[n(z)] + V_{\rm ext} (z)$. If the gas is flowing with a
constant current density $j=n(z)v(z)$, the Bernoulli equation for
the stationary velocity field $v(z)$ is
\begin{equation}
\mu_j = \frac{m}{2} \left[ \frac{j}{n(z)}\right]^2 + \mu(n) +
V_{\rm ext} (z),
\label{eq:mu}
\end{equation}
where $\mu_j$ is the $z$-independent value of the chemical
potential.

In the following two typical cases, the uniform gas has
a polytropic equation of state,
\begin{equation}
\mu(n)=\alpha n^\gamma :
\end{equation}
1) a dilute Bose gas with repulsive interaction, where
$\gamma=1$ and $\alpha=g=4\pi\hbar^2a_s/m$, and
2) a dilute Fermi gas at
unitarity, where $\mu(n)=(1+\beta) E_{\rm F}
=[(1+\beta) (3\pi^2)^{2/3}\hbar^2/(2m)] n^{2/3}$, i.e.,
$\gamma=2/3$ and $\alpha=(1+\beta) (3\pi^2)^{2/3}\hbar^2/(2m)$.
Here, $\beta$ is a universal parameter, which is negative, and its
absolute value is of order unity, 
accounting for the attractive interatomic interactions \cite{giorgini,beta}.

Using the equation of state, one can write 
\begin{equation}
m c_s^2(z) = n \frac{\partial}{\partial n} \mu(n) = \gamma \mu(n) \, ,
\label{eq:mc2}
\end{equation}
where $c_s(z)$ is the local sound velocity, which depends on $z$ 
through the density profile $n(z)$. In a uniform gas of density $n$, 
the sound velocity is given by $c_s^{(0)}= [\gamma\mu(n)/m]^{1/2}$.

\section{EQUATION OF STATE, INCOMPRESSIBILITY, AND EFFECTIVE MASS\label{sec:thermo}}

In this section, focusing on superfluid Fermi gases at unitarity,
we discuss the effects of the periodic potential on 
the macroscopic and the static properties of the fluid, such as 
the equation of state, the incompressibility, and the effective mass \cite{optlatunit}.
The important point is that the periodic potential
favors the formation of bound molecules in a two-component Fermi gas
even at unitarity \cite{orso}
(see also, e.g., Refs.\ \onlinecite{fedichev} and \onlinecite{moritz}).
The emergence of the lattice-induced bound states
drastically changes the above macroscopic and static properties
from those of uniform systems in the strong lattice region of
$\eta_{\rm height}\gg 1$ \cite{optlatunit}.
Such an effect is absent in ideal Fermi gases and 
BECs of repulsively-interacting bosonic atoms, which can be considered
as two limits in the BCS-BEC crossover.

\subsection*{3.1.\quad Basic Equations}

At zero temperature, the chemical potential $\mu$ (the equation of state)
and the current $j$ of a superfluid Fermi gas 
in a lattice are given by the derivatives of the average energy density $\bar{e}=E/V$ 
with respect to the average (coarse-grained) density $\bar{n}$ \cite{note_nbar}
and the average quasi-momentum $\bar{P}$ of the bulk superflow, respectively
(hereafter, for notational simplicity, we omit ``$\bar{\quad}$''
for the coarse-grained quantities, which should not be confused with
the local quantities):
\begin{equation}
  \mu = \frac{ \partial e(n,P)}{ \partial n}\ , \quad
  j = \frac{\partial e(n,P)}{\partial P}\, .
\end{equation}
The incompressibility (or inverse compressibility)
$\kappa^{-1}$ and the effective mass $m^*$
are given by the second derivatives of $e$ with respect to $n$ and $P$:
\begin{align}
  \kappa^{-1} =& n\frac{\partial^2 e(n,P)}{\partial n^2} 
= n \frac{\partial \mu(n,P) }{ \partial n}\, ,\label{eq:kappainv}\\
  \frac{1}{m^*} =& \frac{1}{n} \frac{\partial^2 e(n,P)}{\partial P^2}
= \frac{1}{n}\frac{\partial j(n,P)}{\partial P}\, .
\label{eq:m*}
\end{align}
We calculate these quantities for $P=0$, i.e., for a gas at 
rest, in the periodic potential.

In the absence of the lattice potential ($s=0$), 
the thermodynamic properties of unitary Fermi gases show a universal behavior:
the only relevant length scale is the
interparticle distance fixed by $k_{\rm F}$. Due to translational
invariance, one can write $e(n,P) = e(n,0) + n P^2/2m$ so that
$j=nP/m$ and $m^*=m$. 
Furthermore, the energy density at $P=0$ can be written as
$e(n,0)=(1+\beta)e^0(n,0)$, where 
$e^0(n,0)\equiv (3/5)n E_{\rm F} \propto n^{5/3}$ is
the energy density of the ideal Fermi gas.
Thus, we have $\mu=(1+\beta)E_{\rm F} + P^2/2m$ and 
$\kappa^{-1}=(2/3)(1+\beta) E_{\rm F}$.

\subsection*{3.2.\quad Equation of State and Density Profile}

\begin{figure}[tbp]
\begin{center}\vspace{0.0cm}
\rotatebox{0}{
\resizebox{10cm}{!} 
{\includegraphics{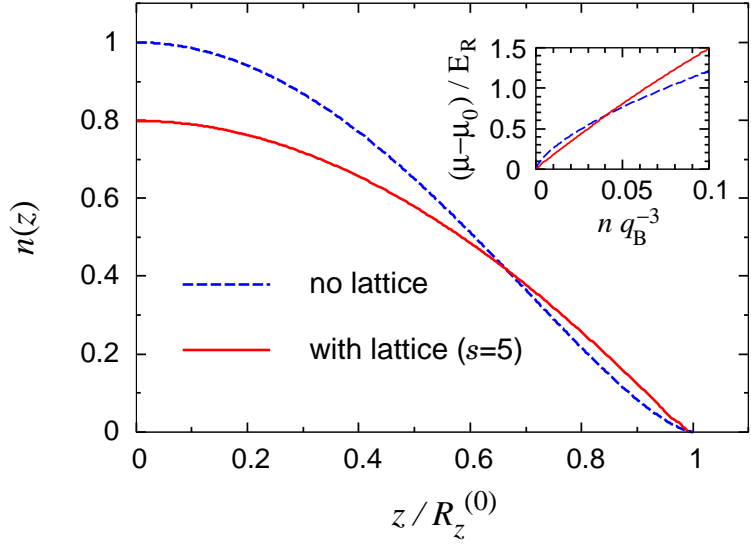}}}
\caption{\label{fig_densprof}(Color online)\quad Coarse-grained
density profiles of a trapped unitary Fermi gas, $n(r_{\perp}=0,z)$ for $s=0$
and $5$ in units of the central density $n(0)=0.0869 q_{\rm B}^3$
calculated for $s=0$ (this local density corresponds to
$E_{\rm F}/E_{\rm R}=1.88$). The quantity $R_z^{(0)}$ is the axial
Thomas-Fermi radius for $s=0$. The inset shows the density
dependence of the chemical potential of unitary Fermi gases. 
Here, $\mu_0$ is the chemical potential in the limit of $n=0$.
This figure is taken from Ref.\ \onlinecite{optlatunit}
}
\end{center}
\end{figure}

When the lattice height $s$ is large, the periodic potential
favors the formation of bound molecules.
In the strong lattice limit $\eta_{\rm height}=s E_{\rm R}/E_{\rm F} \gg 1$,
the system tends to be a BEC of lattice-induced bosonic molecules.
Therefore, in this region, 
the chemical potential shows a linear density dependence, $\mu\propto n$,
as shown by the red solid line in the inset of Fig.\ \ref{fig_densprof}
calculated for unitary Fermi gases in a lattice with $s=5$.
This is clearly different from the density dependence of the 
chemical potential in the uniform system ($s=0$), $\mu\propto n^{2/3}$,
as shown by the blue dashed line in the same inset.
We also note that, for $s \gg 1$, this linear density dependence persists
even at relatively large densities so that $E_{\rm F}/sE_{\rm R}\sim 1$
(e.g., $nq_{\rm B}^{-3}=0.1$ corresponds to $E_{\rm F}/E_{\rm R}\simeq2.1$; thus,
$E_{\rm F}/sE_{\rm R}\simeq 0.41$ in the case of this figure)
because the system effectively behaves like a 2D system in this density region
due to the bandgap in the longitudinal degree of freedom.

This drastic change of EOS manifests itself as a change of
the coarse-grained density profile when a harmonic confinement potential 
$V_{\rm ho}(\mathbf r)$ is added to the periodic potential.
The coarse-grained 
density profile, $n({\mathbf r})$, is calculated using the LDA for $\mu$:
\begin{equation}
  \mu_0 = \mu[n({\mathbf r})] + V_{\rm ho}(\mathbf r) .
\end{equation}
Here, $\mu[n]$ is the local chemical potential as a function of the
coarse-grained density $n$ obtained by using the BdG calculation
for the lattice system
and $\mu_0$ is the chemical potential
of the system fixed by the normalization condition 
$\int d{\mathbf r}\, n({\mathbf r})=N$.
Figure \ref{fig_densprof} clearly
shows that, for $s=5$, the profile takes the form of an inverted
parabola, reflecting the linear density dependence of the chemical
potential (see inset).  In this calculation, we assume an
isotropic harmonic potential,
$V_{\rm ho}(\mathbf r)=m\omega^2r^2/2$,
where $\omega$ is the trapping frequency, 
$\hbar\omega/E_{\rm R}=0.01$, and the number of particles $N=10^6$;
these parameters are close to the experimental ones in Ref.~\onlinecite{miller}.

\subsection*{3.3.\quad Incompressibility and Effective Mass}

\begin{figure}[tbp]
\begin{center}\vspace{0.0cm}
\rotatebox{270}{
\resizebox{!}{16cm}{\includegraphics{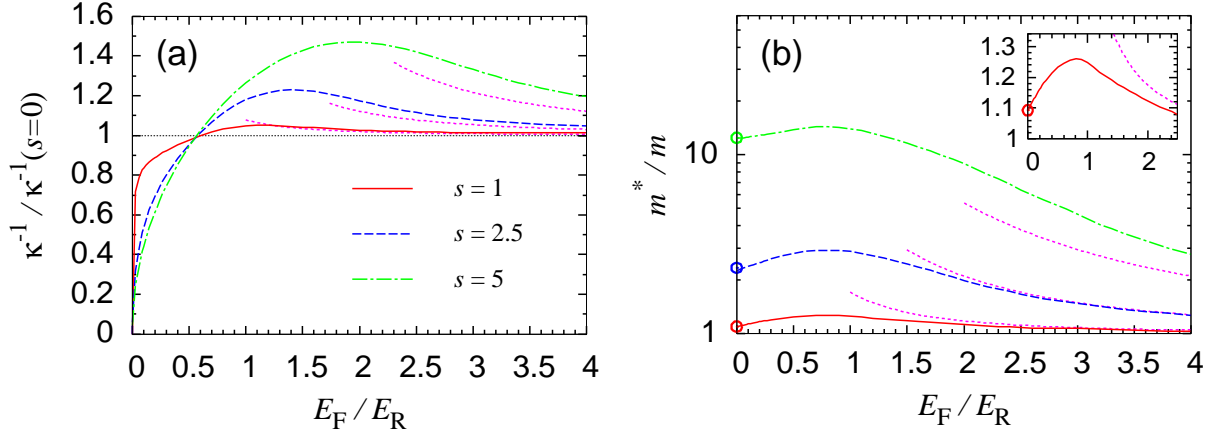}}} 
\caption{\label{fig_kinv_meff_uni}(Color online)\quad 
Incompressibility $\kappa^{-1}$ and effective mass $m^*$ of 
unitary Fermi gases for $s=1$ (red), $2.5$ (blue), and $5$ (green). 
Asymptotic expressions, Eqs.\ (\ref{expansionk-1}) and (\ref{expansionm}), 
obtained by using the hydrodynamic theory are shown by the dotted lines. 
Open circles in panel (b) show $m^*$ from Ref.~\onlinecite{orso}
which was obtained by solving the Schr\"odinger equation for the two-body problem.
The $s=1$ results for $m^*$ are also shown in the inset on the 
linear scale. 
This figure is adapted from Ref.~\onlinecite{optlatunit}.
}
\end{center}
\end{figure}

The formation of molecules induced by the lattice 
also has important consequences for $\kappa^{-1}$ and $m^*$. 
Due to the linear density dependence of the chemical potential 
in the strong lattice region ($\eta_{\rm height}\gg 1$ or $E_{\rm F}/E_{\rm R}\ll s$;
or low density limit for a fixed value of $s$), 
$\kappa^{-1}$ is also proportional to $n$ and
$\kappa^{-1}/\kappa^{-1}(s=0) \propto n^{1/3} \rightarrow 0$
for $E_{\rm F}/E_{\rm R}\rightarrow 0$
[see Fig.\ \ref{fig_kinv_meff_uni}(a)].
This means that the gas becomes highly compressible
in the presence of a strong lattice.
This is in strong contrast to the ideal Fermi gas
corresponding to the BCS limit, which gives nonzero
values of $\kappa^{-1}/\kappa^{-1}(s=0)\sim 1$ 
even in the same limit (see Fig.\ 2 in Ref.\ \onlinecite{optlatunit}).
On the other hand, in the weak-lattice limit (or high-density limit
for a fixed value of $s$), the system reduces to a uniform gas.
By using an hydrodynamic theory, which is valid when $E_{\rm F}/E_{\rm R}\gg 1$, 
and expanding with respect to the
small parameter $sE_{\rm R}/E_{\rm F}$, we obtain $\kappa^{-1}$ 
of unitary Fermi gases in this region as \cite{optlatunit} 
\begin{equation}
\kappa^{-1} \simeq \frac{2}{3}(1+\beta)
E_{\rm F} \left[ 1 +  \frac{1}{32} (1+\beta)^{-2} 
\left(\frac{sE_{\rm R}}{E_{\rm F}}\right)^2 \right]
 + O \bigl[ \left( sE_{\rm R}/E_{\rm F}\right)^4 \bigr] .
\label{expansionk-1}
\end{equation}
This is shown by dotted lines in Fig.~\ref{fig_kinv_meff_uni}(a).
Note that, in this region, $\kappa^{-1}/\kappa^{-1}(s=0)>1$,
and it decreases to unity with increasing $E_{\rm F}/E_{\rm R}$.
Therefore, $\kappa^{-1}/\kappa^{-1}(s=0)$ should 
take a maximum value larger than unity
in the intermediate region of $E_{\rm F}/E_{\rm R}\sim 1$, 
as can be seen in Fig.~\ref{fig_kinv_meff_uni}(a),
which is mainly caused by the bandgap in the longitudinal degree of freedom.

Because the tunneling rate between neighboring sites,
which is related to the (inverse) effective mass $1/m^*$,
is exponentially suppressed with increasing mass,
the formation of molecules induced by the lattice
can yield a drastic enhancement of $m^*$ for $s\gg 1$ in the low-density limit
[Fig.\ \ref{fig_kinv_meff_uni}(b)].
This enhancement makes $m^*$ much larger than it is for ideal Fermi gases
(see Fig.\ 2 in Ref.~\cite{optlatunit})
and for BECs of repulsively-interacting Bose gases 
(see Fig.\ 4 in Ref.~\onlinecite{kramer}) 
with the same mass $m$.
As $E_{\rm F}/E_{\rm R}$ increases, the effective mass exhibits a
maximum at $E_{\rm F}/E_{\rm R}\sim 1$ due to the bandgap in the 
longitudinal degree of freedom; 
then, it decreases to the bare mass, $m^*=m$. 
Hydrodynamic theory can explain 
the behavior of $m^*$ of unitary Fermi gases 
for small $sE_{\rm R}/E_{\rm F}$ \cite{optlatunit}:
\begin{equation}
\frac{m^*}{m} \simeq 1+\frac{9}{32} (1+\beta)^{-2} 
\left(\frac{s E_{\rm R}}{E_{\rm F}}\right)^2 
+ O\bigl[\left(sE_{\rm R}/E_{\rm F}\right)^4\bigr]\ .
\label{expansionm}
\end{equation}
The numerical factor in the second term 
shows that the effect of the lattice is stronger for $m^*$ than for
$\kappa^{-1}$.  It is worth comparing the results with the case
of bosonic atoms, where $m^*$ decreases monotonically with increasing
density because the interaction broadens the condensate wave function
and favors the tunneling \cite{kramer}.

\begin{figure}[tbp]
\begin{center}\vspace{0.0cm}
\rotatebox{0}{
\resizebox{10cm}{!} 
{\includegraphics{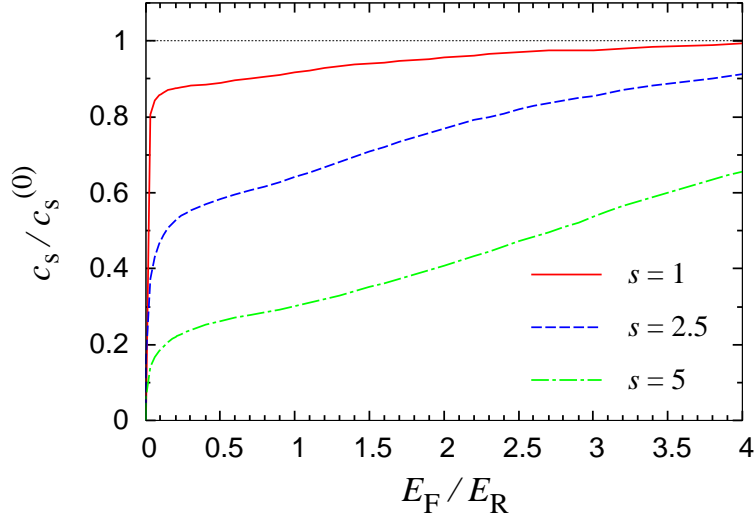}}}
\caption{\label{fig_cs}(Color online)\quad
Sound velocity $c_{\rm s}$ of unitary Fermi gases in a lattice 
in units of the sound velocity
$c_{\rm s}^{(0)} =[(2/3)(1+\beta)E_{\rm F}/m]^{1/2}$
for a uniform system. As in Fig.\ \ref{fig_kinv_meff_uni},
red, blue and green lines correspond to $s=1, 2.5$, and $5$,
respectively.
This figure is taken from Ref.~\onlinecite{optlatunit}
}
\end{center}
\end{figure}

In Fig.~\ref{fig_cs}, we show the sound velocity
of the unitary Fermi gases in a lattice,
\begin{equation}
  c_{\rm s}=\sqrt{\frac{\kappa^{-1}}{m^*}}\, ,
\label{eq:cs}
\end{equation}
calculated from the above results for $\kappa^{-1}$ and $m^*$.
It shows a significant reduction compared to the uniform system,
mainly due to the larger effective mass $m^*/m>1$ except for 
the low-density (more precisely, strong lattice) limit 
in which $\kappa^{-1}$ and, thus, $c_{\rm s}$ show abrupt reductions.
Using Eqs.\ (\ref{expansionk-1}) and (\ref{expansionm}),
we obtain the expression of the sound velocity
in the weak lattice limit as
\begin{equation}
  c_s^2 \simeq {c_s^{(0)}}^2 \left[1-\frac{1}{4}(1+\beta)^{-2}
\left(\frac{s E_{\rm R}}{E_{\rm F}}\right)^2 
+ O\bigl[\left(sE_{\rm R}/E_{\rm F}\right)^4\bigr]\right]\, ,
\end{equation}
where $c_{\rm s}^{(0)} \equiv [(2/3)(1+\beta)E_{\rm F}/m]^{1/2}$
is the sound velocity for a uniform system.

\section{STABILITY\label{sec:stability}}

\subsection*{4.1.\quad Landau Criterion}

Stability of a superfluid flow is one of the most fundamental
issues of superfluidity and was pioneered by Landau \cite{landau}. 
He predicted a critical velocity $v_c$ of the superflow
above which the kinetic energy of the superfluid was 
large enough to be dissipated by creating excitations
(see, e.g., Refs.\ \onlinecite{landau,nozieres_pines,pethick_smith,pitaevskii_stringari}).
This instability is called the Landau or energetic instability, and
its critical velocity is the Landau critical velocity.

The celebrated Landau criterion for energetic instability 
of uniform superfluids is given by \cite{landau,nozieres_pines,pethick_smith,pitaevskii_stringari}
\begin{equation}
  v > v_c = \min{\left(\frac{\epsilon(p)}{p}\right)},
\label{eq:landau}
\end{equation}
where $v$ is the velocity of the superflow,
$\epsilon(p)$ is the excitation spectrum in the static case ($v=0$),
and $p$ is the magnitude of the momentum $\bm{p}$
of an excitation in the comoving frame of the fluid.
Here, $v_c$ is determined by the condition for which
there starts to exist a momentum $\bm{p}$ at which the excitation spectrum
in the comoving frame of a perturber 
(it can be an obstacle moving in the fluid or a vessel in which the fluid
flows) is zero or negative.

For superfluids of weakly interacting Bose gases, the excitation spectrum
is given by the Bogoliubov dispersion relation 
(e.g., Refs.\ \onlinecite{bogoliubov,nozieres_pines,pethick_smith,pitaevskii_stringari})
\begin{equation}
  \epsilon(p)= \sqrt{\frac{p^2}{2m}\left(\frac{p^2}{2m} + 2 gn \right)}
= c_s p \sqrt{1+\left(\frac{p}{2mc_s}\right)^2}\ ,
\end{equation}
with
\begin{equation}
  c_s= \sqrt{\frac{gn}{m}}
\label{eq:csbose}
\end{equation}
being the sound velocity defined by Eq.\ (\ref{eq:cs})
[note that $\mu=gn$ and, thus, $\kappa^{-1}=gn$, and $m^*=m$
for uniform BECs].
Thus, from Eqs.\ (\ref{eq:landau}) and (\ref{eq:csbose}),
the Landau critical velocity
is easily seen to be given by the sound velocity $v_c=c_s$ for $p=0$.
This means that, in superfluids of dilute Bose gases,
the energetic instability is caused by excitations of long-wavelength phonons.
We also note that BECs of non-interacting Bose gases
cannot show superfluidity in a sense that $v_c=0$ and
they cannot support a superflow.

In superfluid Fermi gases, another mechanism 
can cause the energetic instability: fermionic pair-breaking excitations.
In the mean-field BCS theory, the quasiparticle spectrum of
uniform superfluid Fermi gases is given by
\begin{equation}
  \epsilon(p)=\sqrt{\left(\frac{p^2}{2m}-\mu\right)^2+\Delta^2}\ .
\end{equation}
Thus, the Landau critical velocity due to the pair-breaking excitations
is given by \cite{combescot}
\begin{equation}
  v_c = \sqrt{\frac{1}{m}\left(\sqrt{\mu^2+\Delta^2} - \mu\right)}\ .
\end{equation}
In the deep BCS region, where $\mu\simeq E_{\rm F}\gg \Delta$,
we obtain $v_c\simeq \Delta/p_{\rm F}$ with $p_{\rm F}\equiv \hbar k_{\rm F}$.

In the BCS-BEC crossover of superfluid Fermi gases,
where the above two kinds of excitations exist,
the Landau critical velocity is determined by which of them
gives smaller $v_c$.
In the weakly-interacting BCS region ($1/k_{\rm F}a_s\alt -1$), 
the pairing gap is exponentially small, $\Delta \sim E_{\rm F} e^{-\pi/2k_{\rm F}|a_s|}$,
and $v_c$ is set by the pair-breaking excitations.
On the other hand, in the BEC region ($1/k_{\rm F}a_s\agt 1$),
where the system consists of weakly-interacting bosonic molecules, 
creation of long-wavelength superfluid phonon excitations
causes an energetic instability.
In the unitary region, both 
mechanisms are suppressed, and the critical velocity shows a maximum 
value \cite{andrenacci,combescot,sensarma}.

\subsection*{4.2.\quad Stability of Superflow in Lattice Systems}

\subsubsection*{4.2.1.\quad Energetic instability and dynamical instability}

For the onset of an energetic instability, energy dissipation is necessary
in general; i.e., in closed systems, $v>v_c$ is not a sufficient
condition for a breakdown of the superflow, so the flow could still persist
even at $v>v_c$.  The energetic instability corresponds to the situation
in which the system is located at a saddle point of the energy landscape
(i.e., there is at least one direction in which the curvature
of the energy landscape is negative).

In the presence of a periodic potential, another type of
instability, called dynamical (or modulational) instability, can occur
in addition to the energetic instability.
The dynamical instability means that small perturbations on a
stationary state grow exponentially in the process of
(unitary) time evolution without dissipation.
Similar to energetically-unstable states,
dynamically-unstable states are also located at saddle points 
in the energy landscape, but a difference from energetically-unstable states
is that there are kinematically-allowed excitation processes that satisfy
the energy and (quasi-)momentum conservations.
This means that an energetic instability is a necessary condition 
for dynamical instability (see, e.g., Refs.\ \onlinecite{wu2001} and \onlinecite{wu2003}
for bosons and Ref.\ \onlinecite{ring} for fermions); therefore,
the critical value of the (quasi-)momentum for the dynamical instability
should always be larger than (or equal to) that for the energetic instability.
For such a reason, we shall focus on the energetic instability hereafter
\cite{note:dynamical}.

\subsubsection*{4.2.2.\quad Determination of the critical velocity}

If the energetic instability is caused by 
long-wavelength superfluid phonon excitations,
the critical velocity can be determined by using a hydrodynamic analysis
of the excitations 
\cite{machholm03,taylor,pitaevskii05,pethick_smith,vc,vc_crossover}
(this should not be confused with the LDA hydrodynamics
discussed in Sec.\ 2.3).
This analysis is valid provided that the wavelength of the excitations
that trigger the instability is much larger than the 
typical length scale of the density variation, i.e., the lattice constant $d$.

We consider the continuity equation and the Euler equation for the 
coarse-grained density $n$ and the coarse-grained 
(quasi-)momentum $P$ averaged over 
the length scale larger than the lattice constant:
\begin{align}
  \frac{\partial n}{\partial t}+\bm{\nabla}\cdot\bm{j}=&
\frac{\partial n}{\partial t}+\frac{\partial}{\partial z} \frac{\partial e}{\partial P}=0 ,\\
  \frac{\partial P}{\partial t}+ \frac{\partial \mu}{\partial z}=&
\frac{\partial P}{\partial t}+ \frac{\partial}{\partial z} \frac{\partial e}{\partial n}=0 ,
\end{align}
where $e=e(n,P)$ is the energy density of the superfluid 
in the periodic potential for the averaged density $n$ and 
the averaged (quasi-)momentum $P$.
Linearizing with respect to the perturbations of
$n(z,t)=n_0+\delta n(z,t)$ and $P(z,t)=P_0+\delta P(z,t)$
with $\delta n(z,t)$ and $\delta P(z,t)\propto e^{iqz-i\omega t}$,
we obtain the dispersion relation of the long-wavelength phonon,
\begin{equation}
  \omega(q) = \frac{\partial^2 e}{\partial n \partial P} q
+ \sqrt{\frac{\partial^2 e}{\partial n^2}\frac{\partial^2 e}{\partial P^2}}\,
|q|\, .
\label{eq:dispersion}
\end{equation}
Here, $\hbar\omega$ and $q$ are the energy and the wavenumber of the
excitation, respectively. In the first term 
(so-called Doppler term), $\partial_n\partial_P e \ge 0$.
Thus, the energetic instability occurs when 
$\omega(q)$ for $q=-|q|$ becomes zero or negative:
\begin{equation}
\frac{\partial^2 e}{\partial n \partial P} \ge
\sqrt{\frac{\partial^2 e}{\partial n^2}
\frac{\partial^2 e}{\partial P^2}}\ .
\label{eq:vcrit-hydro}
\end{equation}
Using this condition, we determine the critical quasi-momentum $P_c$
at which the equality of Eq.\ (\ref{eq:vcrit-hydro}) holds,
and finally, we obtain the critical velocity from
\begin{equation}
  v_c = \frac{1}{n}\left(\frac{\partial e}{\partial P}\right)_{P_c}\, .
\label{eq:vc}
\end{equation}

We note that, for calculating $P_c$ and $v_c$ using 
Eqs.~(\ref{eq:vcrit-hydro}) and (\ref{eq:vc}), 
what we only need is the energy density of the stationary states 
as a function of $n$ and $P$.
This can be obtained by solving, e.g., the GP or the BdG equations
for the periodic potential.

If the energetic instability of Fermi superfluids
is caused by pair-breaking excitations,
the critical velocity can be determined by
using the quasiparticle energy spectrum $\epsilon_i$ obtained 
from the BdG equations.
The energetic instability due to the pair-breaking excitations occurs
when some quasiparticle energy $\epsilon_i$ becomes zero or negative:
\begin{equation}
  \epsilon_i\le 0.
\end{equation}
We obtain a critical velocity for the pair-breaking excitations
from Eq.\ (\ref{eq:vc}) evaluated at the critical quasi-momentum
determined by this condition.

\subsubsection*{4.2.3.\quad Critical velocity of 
superfluid Bose gases and superfluid unitary Fermi gases in a lattice}

First, we consider the situation where the LDA is valid; i.e.,
the lattice constant $d$ is much larger than the healing length $\xi$.
This condition corresponds to $gn/E_{\rm R}\gg 1$ for superfluid Bose gases
and $E_{\rm F}/E_{\rm R}\gg 1$ for superfluid Fermi gases at unitary
(see discussion in Sec.\ 2.1).
In the framework of the LDA hydrodynamics explained in Sec.\ 2.3, 
the system is considered to become energetically unstable
if there exists some point $z$ at which the local superfluid velocity
$v(z)$ is equal to or larger than the local sound velocity $c_s(z)$.
If the external potential is assumed to have a maximum at $z=z_0$ 
[i.e., $V(z_0)=V_0$ for our periodic potential],
then at the same point, the density is minimum, 
$c_s(z)$ is minimum, and $v(z)$ is maximum due to the current conservation;
i.e., $j=n(z)v(z)$ being constant.
This means that the 
superfluid first becomes unstable at $z=z_0$. Using Eq.\ (\ref{eq:mc2}), we can write the 
condition for the occurrence of the instability as 
$m[j_c/n_c(z_0)]^2=\gamma \mu[n_c(z_0)]= \gamma \alpha n_c^\gamma(z_0)$, 
where $n_c(z)$ is the density profile calculated at the critical 
current \cite{kink}. By inserting this condition into Eq.~(\ref{eq:mu}),
we can obtain the following implicit relation for the critical current
\cite{vc}:
\begin{equation}
j_c^2 = \frac{\gamma}{m \alpha^{2/\gamma}}
\left[ \frac{2\mu_{j_c}}{2+\gamma}
\left(1-\frac{V(z_0)}{\mu_{j_c}}\right)
\right]^{\frac{2}{\gamma}+1} \, .
\label{eq:jc}
\end{equation}
It is worth noticing that this equation contains only $z$-independent
quantities. It is also independent of the shape of the external
potential: the only relevant parameter is its maximum value
$V(z_0)$. Moreover, it can be applied to both Bose gases and
unitary Fermi gases
(see also Refs.~\onlinecite{mamaladze,hakim,leboeuf} for bosons).

\begin{figure}[tbp]
\begin{center}\vspace{0.0cm}
  \rotatebox{0}{ \resizebox{10cm}{!} 
     {\includegraphics{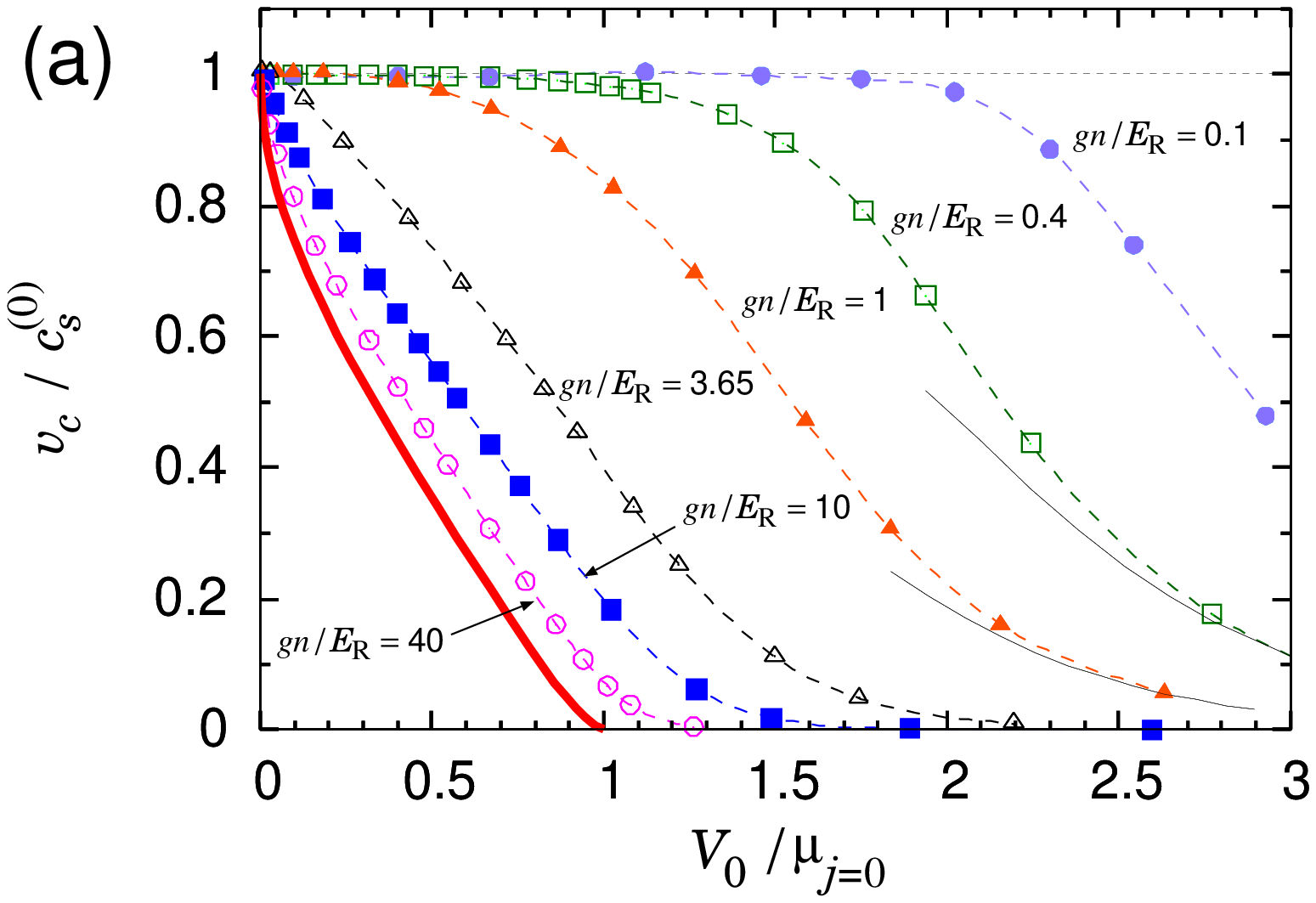}}}\vspace{0.5cm}
   \rotatebox{0}{\resizebox{10cm}{!} 
    {\includegraphics{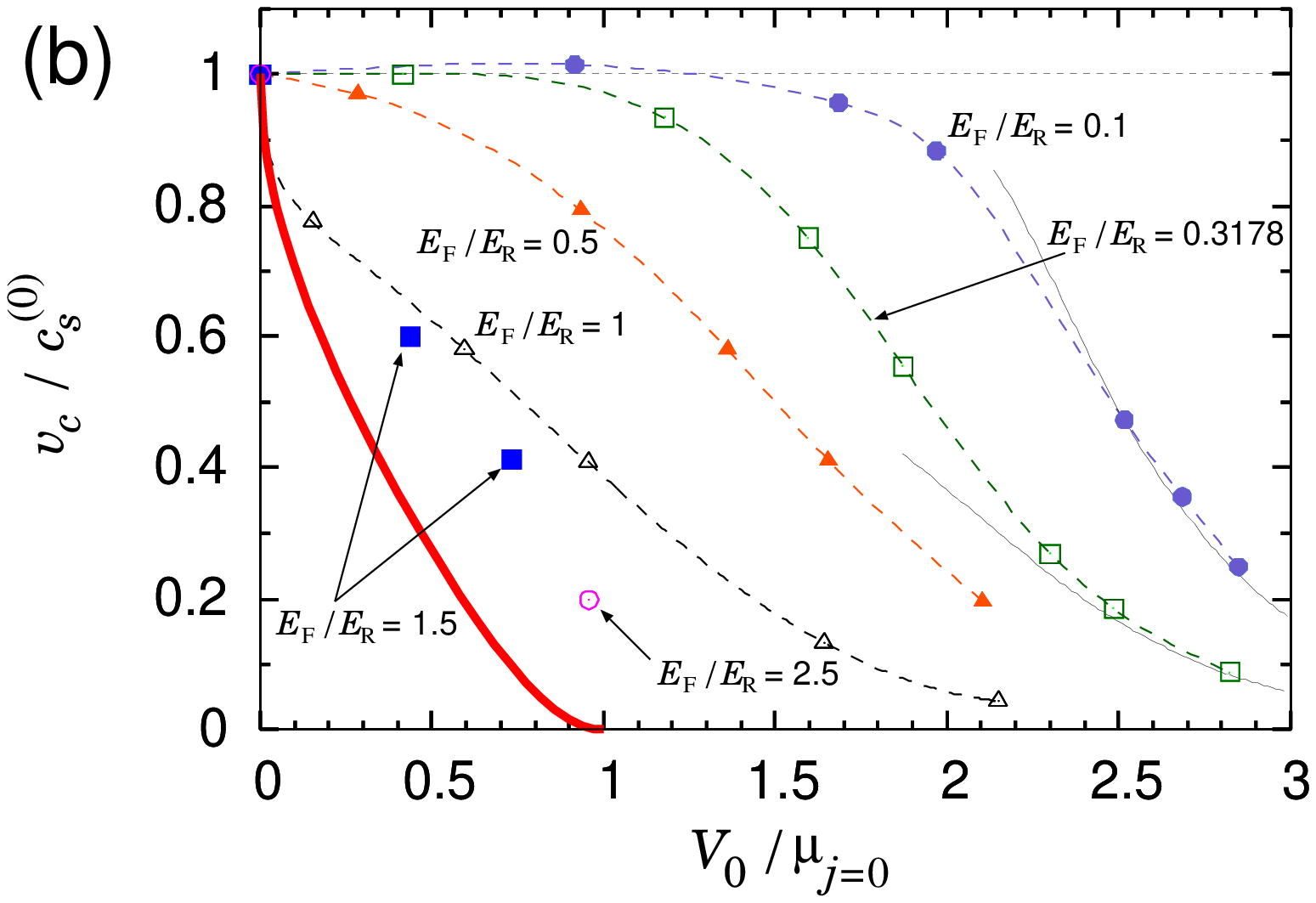}}}
 \caption{\label{fig:vc}(Color online)\quad Critical velocity $v_c$
for energetic instability of superfluids in a 1D periodic potential.
Panel (a) is for superfluids of dilute Bose gases, 
and panel (b) is for superfluids of dilute Fermi gases at unitarity.
The critical velocity is given in units
of the sound velocity of a uniform gas, $c_s^{(0)}$,  and is plotted
as a function of the maximum of the external potential in units of the
chemical potential $\mu_{j=0}$ of the superfluid at rest. 
Thick solid lines: prediction of the hydrodynamic
theory within the LDA, as calculated from Eq.~(\ref{eq:jc}). 
Symbols:
results obtained from the numerical solutions of the GP equation
[panel (a)] and the BdG equations [panel (b)].
The thinner black solid lines are the tight-binding prediction,
Eq.\ (\protect\ref{eq:vctb}).
Dashed lines are guides for the eye.
This figure is adapted from Ref.~\onlinecite{vc}.
}
\end{center}
\end{figure}

In Fig.~\ref{fig:vc}, we plot as thick solid red lines the critical velocity in a lattice
obtained from the hydrodynamic expression in Eq.\ (\ref{eq:jc})
for BECs [panel (a)] and unitary Fermi gases [panel (b)].
In both cases, the critical velocity
$v_c=j_c/n_0$ ($n_0$ is the average density)
is normalized to the value of the sound velocity in
the uniform gas, $c_s^{(0)}$, and is plotted as a function of
$V_0/\mu_{j=0}$. 
The limit $V_0/\mu_{j=0} \to 0$ corresponds to the usual Landau 
criterion for a uniform superfluid flow in the presence of a small 
external perturbation, i.e., a critical velocity equal to the sound
velocity of the gas. In this hydrodynamic scheme, as mentioned before,
the critical velocity decreases when $V_0$ increases
mainly because the density has a local depletion and the velocity
has a corresponding local maximum, so that the Landau instability
occurs earlier. When $V_0 = \mu_{j_c}$, the density
exactly vanishes at $z=z_0$; hence, the critical velocity
goes to zero.

Next, we discuss the critical velocity of the energetic instability
beyond the LDA \cite{vc}.
Here, we use the energy density $e(n,P)$ of superfluids 
in a periodic potential 
calculated by using the mean-field theory in the continuum model,
i.e., the GP equation for bosons and the BdG equations for fermions.
Based on the energy density,
we determine the critical quasi-momentum and the critical velocity
from Eqs.~(\ref{eq:vcrit-hydro}) and (\ref{eq:vc}).

For Bose superfluids, we plot $v_c$ for various values of $gn/E_{\rm R}$
in Fig.\ \ref{fig:vc}(a).
We can clearly see that these results 
approach the LDA prediction for $gn/E_{\rm R}\gg 1$, as expected.
We also note that $v_c$
exhibits a plateau for $gn/E_{\rm R} \alt 1$ (i.e., $\xi\agt d$) and small $V_0$.
This can be understood as follows:
If the healing length $\xi$ is larger than the lattice spacing $d$
and $V_0/gn$ is not too large, the energy 
associated with quantum pressure, which is proportional to $1/\xi^2$,  
acts against local deformations of the order parameter, and the latter 
remains almost unaffected by the modulation of the external 
potential. This is the region of the plateau in Fig.~\ref{fig:vc}.
In terms of Eq.~(\ref{eq:vcrit-hydro}), this region occurs when the 
left-hand side is $\simeq P/m$ and the right-hand side 
is $\simeq c_s^{(0)}$, so that the critical quasi-momentum obeys 
the relation $P_c/m=c_s^{(0)}$, which is the usual Landau criterion 
for a uniform superfluid in the presence of small perturbers.
With increasing $V_0$, this region ends when $\mu \sim E_{\rm R}$.

If we further increase $V_0$, the
chemical potential $\mu$ becomes larger than $E_{\rm R}$, 
the density is forced to oscillate, and $v_c/c_s^{(0)}$ starts to decrease. 
The system eventually reaches a region of weakly-coupled superfluids 
separated by strong barriers, which is well described by the
tight-binding approximation (also for $gn/E_{\rm R}\agt 1$,
the system enters this region when $V_0$ is sufficiently large).
There, the energy density is given by a sinusoidal form 
with respect to $P$ as
\begin{equation}
  e(n,P) = e(n,0) + \delta_J \left[1-\cos{(\pi P/P_{\rm edge})}\right]\, . 
\end{equation}
Here, $P_{\rm edge}$ is the
quasi-momentum at the edge of the first Brillouin zone; i.e., 
$P_{\rm edge}=\hbar q_{\rm B}$ for superfluids of bosonic atoms 
($P_{\rm edge}=\hbar q_{\rm B}/2$ for those of fermionic atoms). The
quantity $\delta_J = n P_{\rm edge}^2/\pi^2 m^*$ corresponds to
the half width of the lowest Bloch band. 
Because of the sinusoidal shape of the energy density and 
the large effective mass in the tight-binding limit, 
the critical quasi-momentum is around $P_{\rm edge}/2$.
Thus, we see that, from Eq.~(\ref{eq:vc}), 
the critical velocity is determined by the effective mass as 
\begin{equation}
v_c \simeq \frac{1}{\pi}\frac{P_{\rm edge}}{m^*}\, .
\label{eq:vctb}
\end{equation}
The values of $v_c$
obtained from Eq.~(\ref{eq:vctb}), with $m^*$ extracted from the GP 
calculation of $e(n,P)$, are plotted by thinner black solid lines in Fig.~\ref{fig:vc}(a) 
for $gn/E_{\rm R}=0.4$ and $1$ in the region of $V_0/\mu_{j=0} \agt 2$.

We also calculate the critical velocity by using another method based on a
complete linear stability analysis for the GP energy functional
as in Refs.\ \onlinecite{wu2001,machholm03}, and \onlinecite{modugno}
(see also Refs.\ \onlinecite{pethick_smith} and \onlinecite{wu2003}).  We have checked that the results
agree with those obtained from Eq.~(\ref{eq:vcrit-hydro}) 
based on the hydrodynamic analysis 
to within 1\% over the whole range of $gn/E_{\rm R}$ and $V_0$
considered in the present work.
This confirms that the energetic
instability in the periodic potential is triggered by long-wavelength
excitations, because the excitation energy of the sound mode is the smallest 
in this limit.

For superfluid unitary Fermi gases, we plot $v_c$ for various
values of $E_{\rm F}/E_{\rm R}$ in Fig.~\ref{fig:vc}(b).
We observe qualitatively similar results compared to those for bosons
plotted in Fig.~\ref{fig:vc}(a).
For $E_{\rm F}/E_{\rm R}\alt 1$, the critical velocity $v_c$ 
shows a plateau at small $V_0$,
and it decreases from $\simeq c_s^{(0)}$ with increasing $V_0$.
For larger $E_{\rm F}/E_{\rm R}> 1$, $v_c$ approaches the LDA result 
(thick solid red line).
In the region of large $V_0$ such that $V_0/\mu$ is sufficiently large,
$v_c$ is well described by the tight-binding expression in Eq.\ (\ref{eq:vctb})
plotted by thinner black solid lines in Fig.~\ref{fig:vc}(b).
Here, we use $m^*$ calculated from the BdG equations.
We also note that the pair-breaking excitations
are irrelevant to the energetic instability at unitarity
except for very low densities such that $E_{\rm F}/E_{\rm R}\ll 1$ and
small, but nonzero, $V_0$.
These excitations are important on the BCS side of the
BCS-BEC crossover, as discussed in Sec.\ 4.2.4.

Finally, we would like to point out that, in the LDA limit, while
the quantities discussed in Sec.\ \ref{sec:thermo}
approach the value in the uniform system,
the critical velocity of the energetic instability is most strongly affected
by the lattice.
Figure \ref{fig:vc} clearly shows that, as we increase the lattice height, 
the critical velocity decreases from that in the uniform system most rapidly 
in the LDA limit.
The main reason for this opposite tendency is that
the critical velocity is determined only around the potential maxima
while the quantities discussed in Sec.\ \ref{sec:thermo}
are determined by contributions from the whole region of the system.

\subsubsection*{4.2.4.\quad Along the BCS-BEC crossover\label{sec:vc_crossover}}

\begin{figure}[tbp]
\begin{center}\vspace{0.0cm}
\rotatebox{0}{
\resizebox{7.7cm}{!}
{\includegraphics{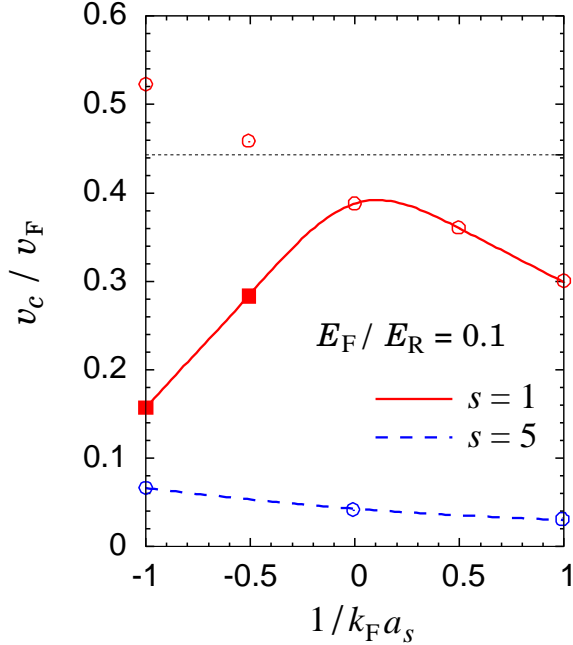}}}
\caption{\label{fig:vc_bcsbec}(Color online)\quad
Critical velocity $v_c$ of the energetic instability
for $E_{\rm F}/E_{\rm R}=0.1$ with $s=1$ and $5$ in the BCS-BEC crossover.
Open circles and filled squares show the critical velocity 
due to long-wavelength phonons and fermionic pair-breaking excitations,
respectively.
The horizontal dotted line represents the 
value of the sound velocity
$c_s^{(0)}$ of a uniform system at unitarity, 
$c_s^{(0)}/v_{\rm F}=(1+\beta)^{1/2}/\sqrt{3}\simeq 0.443$.
The red solid line is a guide for the eye.
}
\end{center}
\end{figure}

Here, we extend our discussion on the Landau critical velocity 
to the BCS-BEC crossover region \cite{vc_crossover}.
As in the uniform systems, both long-wavelength phonon excitations
and pair-breaking excitations can be relevant to
the energetic instability, depending on the interaction parameter
$1/k_{\rm F}a_s$.
However, there is an additional effect due to the lattice:
when the lattice height is much larger than the Fermi energy,
the periodic potential can cause pairs of atoms to be strongly bound
even in the BCS region, so the pair-breaking excitations are suppressed
\cite{vc_crossover}.

In Fig.\ \ref{fig:vc_bcsbec}, we plot $v_c$ 
for superfluid Fermi gases in the BCS-BEC crossover
for different values of the lattice height with $s=1$ and $5$.
Here, we set $E_{\rm F}/E_{\rm R}=0.1$ as an example.
The open circles show $v_c$ of the energetic instability caused by 
long-wavelength superfluid phonon excitations,
and the filled squares show that caused by pair-breaking excitations.
For a moderate lattice strength of $s=1$,
the result for $v_c$ is qualitatively the same as that of the uniform system.
On the BCS side, the smallest $v_c$ is given by the pair-breaking excitations
while, on the BEC side, $v_c$ is set by the long-wavelength phonon excitations,
and around unitarity $v_c$ shows a maximum.
For a larger value of $s=5$, however, the result is very different
from the uniform case.
Due to the lattice-induced binding, pair breaking is suppressed
so that it does not cause an energetic instability
even on the BCS side. 
(Note that there is no filled square for $s=5$.
This means that we do not have a negative quasiparticle energy
for any value of $P$ in the whole Brillouin zone.
This point will be discussed later.)
Throughout the calculated region of $-1\le 1/k_{\rm F}a_s\le 1$,
$v_c$ is determined only by the long-wavelength phonon excitations,
and $v_c$ decreases monotonically with increasing $1/k_{\rm F}a_s$.

\begin{figure}[tbp]
\begin{center}\vspace{0.0cm}
\rotatebox{0}{
\resizebox{8.2cm}{!}
{\includegraphics{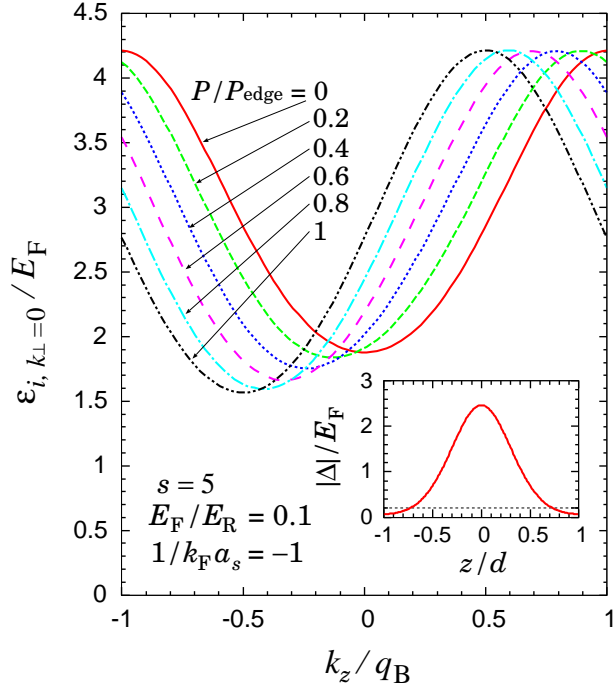}}}
\caption{\label{fig:qp_energy2}(Color online)\quad
Lowest band of the quasiparticle energy spectrum $\epsilon_i$ 
for large lattice height with $s=5$ and $E_{\rm F}/E_{\rm R}=0.1$
(i.e., $E_{\rm F}/V_0=0.02$)
in the BCS region at $1/k_{\rm F}a_s=-1$.
Here, we show the first radial branch with 
$k_\perp^2 \equiv k_x^2+k_y^2=0$, which always gives the smallest values
of $\epsilon_i$ in this case.
The inset shows the amplitude $|\Delta(z)|$ of the order parameter
at $P=0$. The horizontal dotted line shows the amplitude of the order 
parameter for a uniform system at the same value of $1/k_{\rm F}a_s=-1$.
This figure is adapted from Ref.\ \onlinecite{vc_crossover}.
}
\end{center}
\end{figure}

The effect of the lattice-induced molecular formation can be clearly
seen in the quasiparticle energy spectrum $\epsilon_i$ and in the enhancement
of the order parameter $\Delta(z)$. In Fig.\ \ref{fig:qp_energy2},
we show $\epsilon_i$ and $|\Delta(z)|$
in the case of $s=5$, $E_{\rm F}/E_{\rm R}=0.1$, and $1/k_{\rm F}a_s=-1$.
The spectrum for $P=0$ shows a quadratic 
dependence on $k_z$ with a positive curvature around $k_z=0$, and 
there are no minima at $k_z\ne 0$.
Even though the figure represents 
a case in the BCS region, the structure of $\epsilon_i$ is 
consistent with the formation of bound pairs.
We can also see that the spectrum never becomes negative for
any values of $P$ in the first Brillouin zone.
In the inset of the 
same figure, we show the amplitude $|\Delta(z)|$ of the order parameter 
at $P=0$. 
Here, we see a large enhancement of $|\Delta(z)|$ near $z=0$ compared to 
the uniform system, which shows the formation of bosonic bound 
molecules.
We also note that the minimum value of 
$|\Delta(z)|$ at $z/d=\pm 1$ is smaller than, but still comparable 
to, the value of $|\Delta|$ in the uniform case, suggesting that 
the system is, indeed, in the superfluid phase.

\subsubsection*{4.2.5.\quad Experiments}

In closing this section, we briefly summarize experimental
studies on the stability of a superfluid.
Using cold atomic gases, stability of the superfluid
and its critical velocity was first experimentally studied in Ref.\ \onlinecite{raman}
and further examined in Ref.\ \onlinecite{onofrio}.
These experiments used a large (diameter $\gg \xi$) and strong
(height $\gg \mu$) vibrating circular potential in a BEC.
However, it has been concluded that what was observed
in these works was not likely to be the energetic instability:
dynamical nucleation of vortices by vibrating potential \cite{onofrio}.
Recently, Ramanathan {\it et al.} performed a new experiment
on the stability of a superflow
with a different setup \cite{ramanathan}. 
They used a BEC flowing in a toroidal trap
with a tunable weak link (width of the barrier along the flow direction
 much larger than $\xi$).
In that experiment, they obtained a critical velocity
consistent with the energetic instability due to vortex excitations
\cite{feynman}.
For 2D Bose gases, the stability of the superfluidity and 
the critical velocity were also studied recently \cite{desbuquois}.

Regarding the superflow in optical lattices, its stability
was first experimentally studied in Ref.\ \onlinecite{burger}.
In that experiment, they used a cigar-shaped BEC that underwent 
a center-of-mass oscillation in a harmonic trap
in the presence of a weak 1D optical lattice, and
they measured the critical velocity.
Although their original conclusion was that they 
had measured the Landau critical velocity for the energetic instability,
further careful experimental \cite{desarlo} and 
theoretical \cite{wu2001,modugno} follow-up studies 
clarified that the instability observed in Ref.\ \onlinecite{burger}
was a dynamical instability.
In this follow-up experiment by De Sarlo {\it et al.} \cite{desarlo}, 
they employed an improved
setup, i.e., a 1D optical lattice moving at constant 
and tunable velocities instead of an oscillating BEC in a static lattice.
With this new setup, they succeeded in observing
both energetic \cite{desarlo} and 
dynamical \cite{desarlo,fallani} instabilities,
and obtained a very good agreement with theoretical predictions 
\cite{desarlo,modugno,fallani}.
It is worthwhile to stress that this understanding was finally obtained through
a continuous effort over a few years by the same group.

Experimental study on the stability of superfluid Fermi gases
in optical lattices was carried out by Miller {\it et al.} \cite{miller}.
Similar to the experiment of Ref.\ \onlinecite{desarlo}, they also used 
a 1D lattice moving at constant and tunable velocities.
A different point is that, instead of imposing a periodic potential
on the whole cloud, they produced a lattice potential 
only in the central region of the cloud.
They measured the critical velocity of the energetic instability
in the BCS-BEC crossover and found that it was largest
around unitarity.
They also did a systematical measurement of the critical velocity at unitarity
for various lattice heights. However, there is a significant
discrepancy from a theoretical prediction \cite{vc},
so further studies are needed.

\section{ENERGY BAND STRUCTURE \lasec{energy_band}}

In this section, we discuss how the superfluidity can affect 
the energy band structures of ultracold atomic gases in periodic optical potentials.
Starting with the noninteracting particles in a periodic potential, 
we obtain the well-known sinusoidal energy band structure. 
When the interparticle interaction is turned on in such a way that the superfluidity appears, 
we see a drastic change in the energy band structure.
For BECs, it has been pointed out that the interaction 
can change the Bloch band structure, causing the appearance of 
a loop structure called ``swallowtail'' in the energy dispersion \cite{wu_st,diakonov,seaman,danshita}. 
This is due to the competition between the external periodic potential and the nonlinear mean-field interaction: the former favors a sinusoidal band structure while the latter tends to 
make the density smoother and the energy dispersion quadratic. When the nonlinearity wins, the effect of the external potential is screened, and a swallowtail energy loop appears \cite{mueller}. This nonlinear effect requires the existence of an order parameter; consequently, the 
emergence of swallowtails can be viewed as a peculiar manifestation of superfluidity in periodic potentials.

Qualitatively, one can argue in a similar way on the existence of the
swallowtail energy band structure in Fermi superfluids in optical
lattices; the competition between the external periodic potential
energy and the nonlinear interaction energy ($g |\sum_i u_i(\mathbf r)
v_i^*(\mathbf r)|^2$) determines the energy band structure.  However,
the interaction energy in the Fermi gas is more involved, and the
unified view along the crossover from the BCS to the BEC states 
is a nontrivial and interesting problem in itself.  To answer the questions of
1) whether or not swallowtails exist in Fermi superfluids and 2)
whether unique features that are different from
those in bosons exist, we solve the BdG equations [see \refeq{BdG2}] of a
two-component unpolarized dilute Fermi gas subject to a
one-dimensional (1D) optical lattice \cite{swallowtail}.

\subsection*{5.1.\quad Swallowtail Energy Spectrum}

\begin{figure}[!tb,floatfix]
\centering
\resizebox{7.25cm}{!} 
{\includegraphics{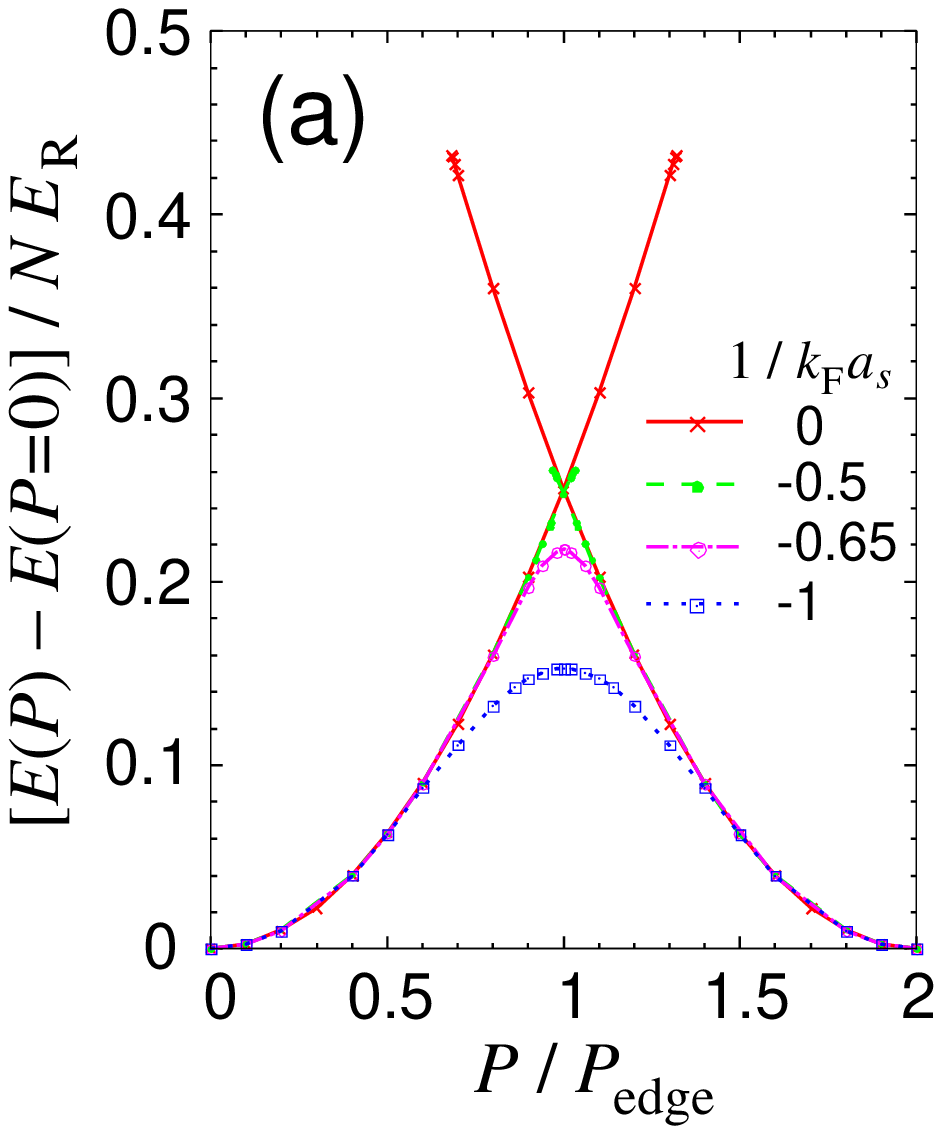}}
\resizebox{7.25cm}{!} 
{\includegraphics{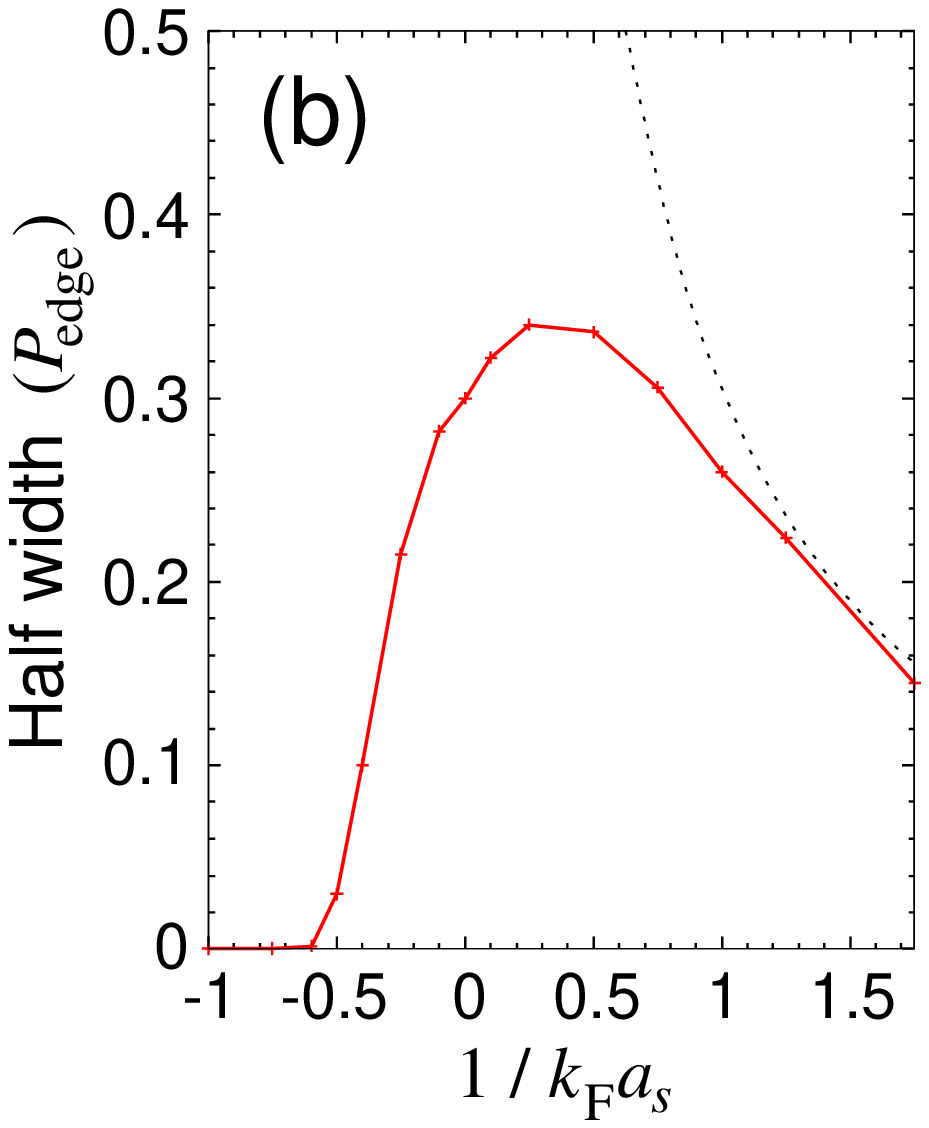}}
\caption{(Color online) (a) Energy $E$ per particle as a function of the
  quasi-momentum $P$ for various values of $1/k_{\rm F}a_s$, and 
  (b) half width of the swallowtails along the BCS-BEC crossover. 
   These results are obtained for $s=0.1$ and $E_{\rm F}/E_{\rm R} 
  = 2.5$. The quasi-momentum $P_{\rm edge}=\hbar q_{\rm B}/2$ fixes
  the edge of the first Brillouin zone. The dotted 
  line in (b) is the 
  half width in a BEC obtained by solving the GP equation; it vanishes 
  at $1/k_{\rm F}a_s \simeq 10.6$.
  This figure is taken from Ref.~\onlinecite{swallowtail}.} \fig{swtail}
\end{figure}

The energy per particle in the lowest Bloch band as a function 
of the quasi-momentum $P$ for various values of $1/k_{\rm F}a_s$ 
is computed \cite{footnote_parameter_value}.
The results in \reffig{swtail}(a) show that the swallowtails appear above a
critical value of $1/k_{\rm F}a_s$ where the interaction energy is strong 
enough to dominate the lattice potential. In \reffig{swtail}(b), the 
half-width of the swallowtails from the BCS to the BEC side is shown.
It reaches a maximum near unitarity ($1/k_{\rm F}a_s=0$).
In the far BCS and BEC limits, the width vanishes because the 
system is very weakly interacting and the band structure tends to 
be sinusoidal.
When approaching unitarity from either side, the interaction energy 
increases and can dominate over the periodic potential, which means that 
the system behaves more like a translationally-invariant superfluid 
and the band structure follows a quadratic dispersion terminating
at a maximum $P$ larger than $P_{\rm edge}$.

\begin{figure}[tb,floatfix]
\centering
\resizebox{8cm}{!} 
{\includegraphics{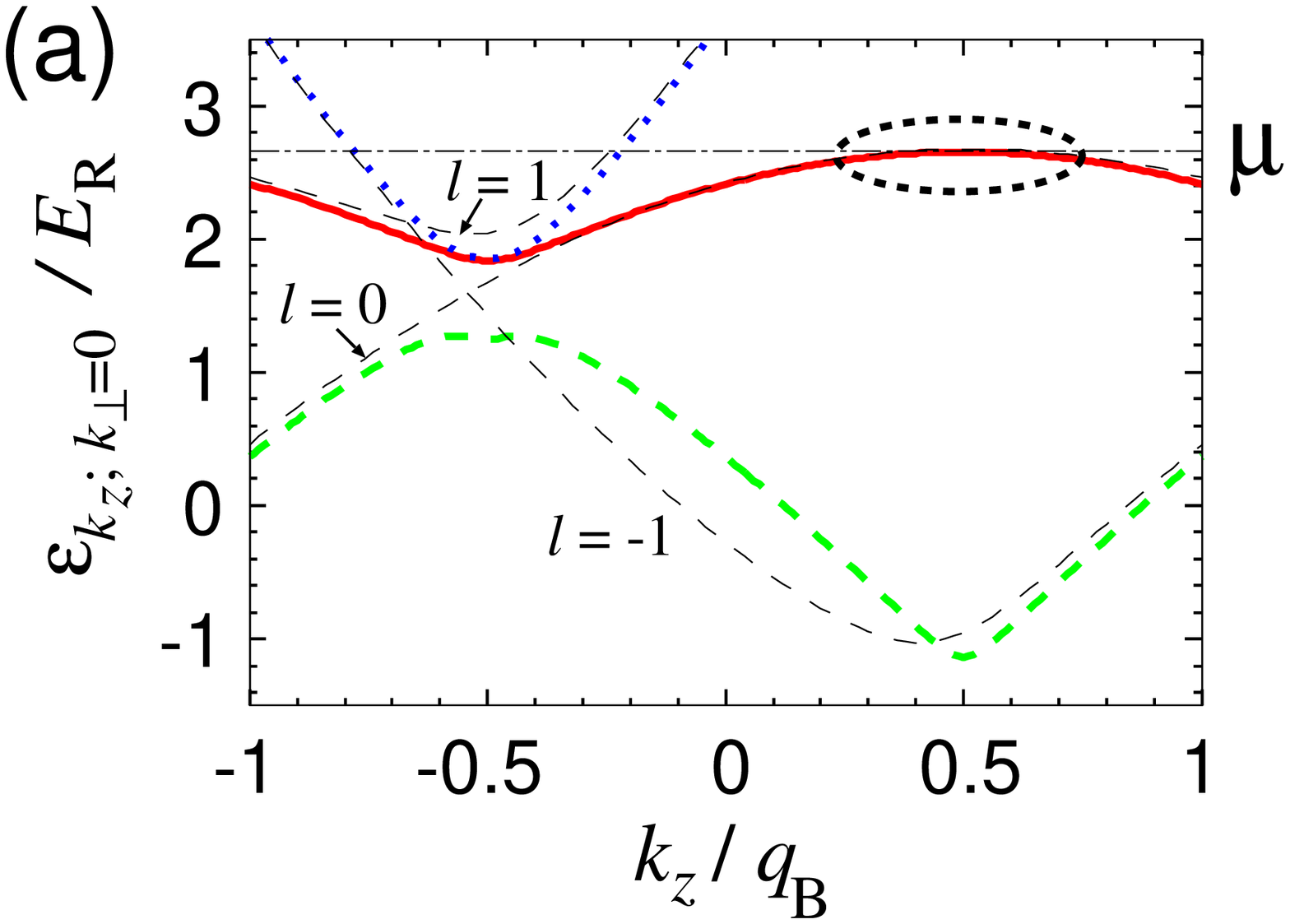}}
\resizebox{8cm}{!} 
{\includegraphics{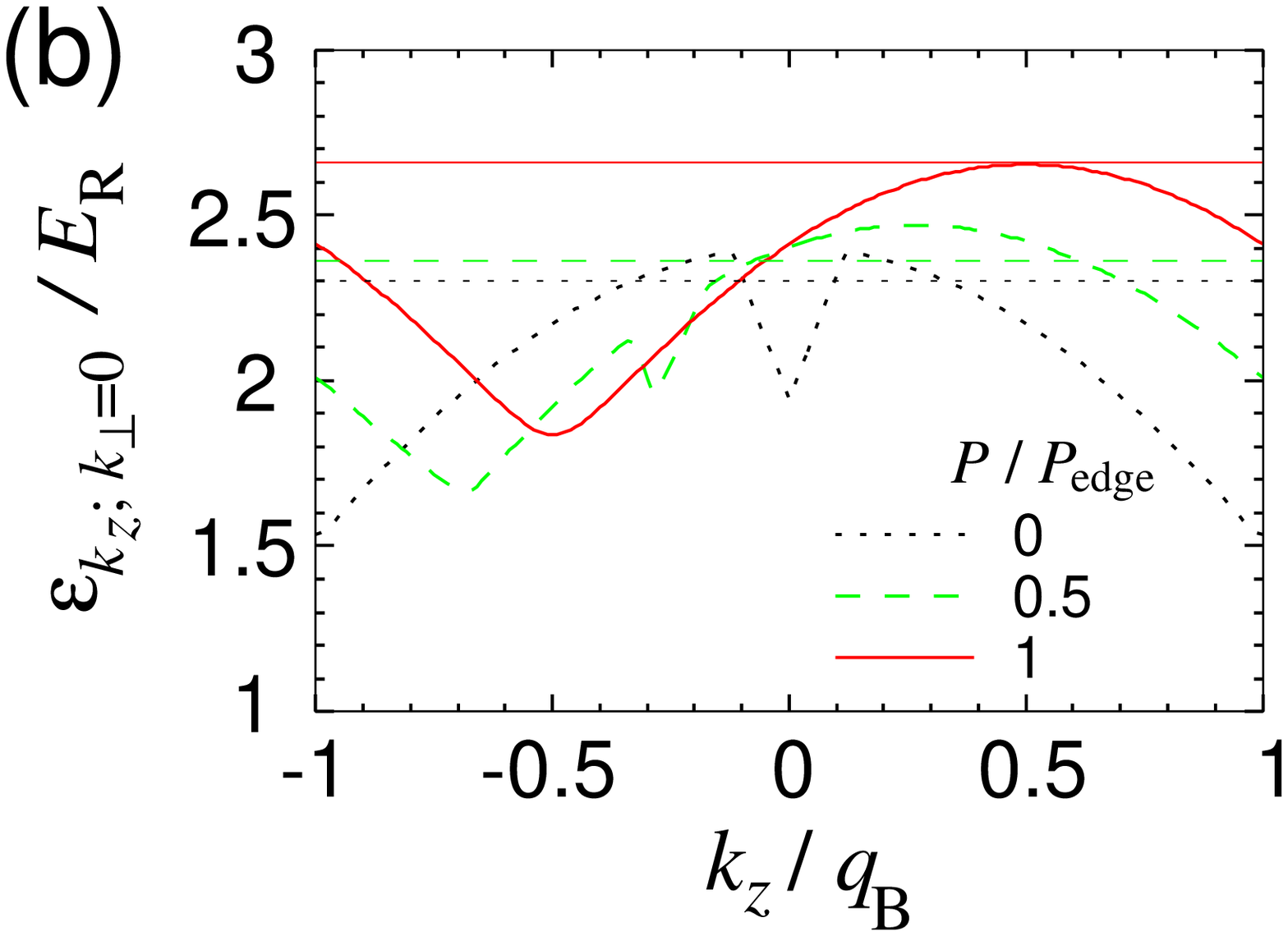}}
\caption{(Color online) (a) Lowest three Bloch bands of the quasiparticle 
energy spectrum at $k_{\perp}=0$ for $P=P_{\rm{edge}}$ 
and $1/k_{\rm F}a_s = -0.62$. 
Thin black dashed lines labeled by $l$'s
show the approximate energy bands obtained from
Eq.~(\ref{band_approx}) by using $\mu \simeq 2.66 E_{\rm R}$
and $|\Delta|\simeq |\Delta(0)|\simeq 0.54 E_{\rm R}$.
(b) Bloch band of the quasiparticle energy spectrum around the
chemical potential for $P/P_{\rm{edge}}=0$ (black dotted), $0.5$
(green dashed), and $1$ (red solid) at $k_{\perp}=0$
and $1/k_{\rm F}a_s = -0.62$ in the case of $s=0.1$ and 
$E_{\rm F}/E_{\rm R} =2.5$. Each horizontal line denotes the value of
the chemical potential for the corresponding value of $P$.
This figure is adapted from Ref.~\onlinecite{swallowtail}.
}
\fig{qp_spectrum}
\end{figure}

The emergence of swallowtails on the BCS side for $E_{\rm F}/E_{\rm R}\agt 1$ is 
associated with peculiar structures of the quasiparticle energy spectrum 
around the chemical potential. 
In the presence of a superflow moving in the $z$ direction with wavevector $Q$ ($\equiv P/\hbar$), 
the quasiparticle energies are given by the eigenvalues 
in \refeq{BdG2}. Because the potential is shallow ($s \ll 1$), some qualitative 
results can be obtained even when ignoring $V_{\text{ext}}(z)$ except for its 
periodicity. With this assumption, we obtain
\begin{equation}
\epsilon_{\mathbf k} \! \approx \! \frac{(k_z \! + \! 2q_{\rm B} l)Q}{m} 
\! + \! \sqrt{\left[ \! \frac{k_\perp^2 \! + \! (k_z \! + \! 2q_{\rm B} l)^2 \! + \! Q^2}{2m} 
\! - \! \mu \right]^2 \!\!\! + \! |\Delta|^2}  \ ,
\label{band_approx}
\end{equation}
with $l$ being integers for the band index. If $Q=0$, the $l=0$ band 
has the energy spectrum $\sqrt{[(k_\perp^2+ k_z^2)/2m - \mu]^2+ |\Delta|^2}$, 
which has a local maximum at $k_z=k_{\perp}=0$. When 
$Q=P_{\rm edge}/\hbar$, the spectrum is tilted, and the local maximum moves 
to $k_z\simeq q_{\rm B}/2$ provided $|\Delta| \ll E_{\rm F}$ (and 
$E_{\rm F}/E_{\rm R}\agt 1$). 
In the absence of the swallowtail, the full BdG calculation, indeed, gives
a local maximum at $k_z=q_{\rm B}/2$, and the quasiparticle spectrum is
symmetric about this point, which 
reflects that the current is zero.
As $E_{\rm F}/E_{\rm R}$ increases, the band becomes flatter
as a function of $k_z$ and narrower in energy.

In \reffig{qp_spectrum}(a), we show the quasiparticle energy spectrum
at $k_{\perp}=0$ for $P=P_{\rm{edge}}$.  When the swallowtail is on the
edge of appearing, the top of the narrow band just
touches the chemical potential $\mu$ [see the dotted ellipse in
\reffig{qp_spectrum}(a)]. Suppose $1/k_{\rm F}a_s$ is slightly larger
than the critical value so that the top of the band is slightly above
$\mu$. In this situation, a small change in the quasi-momentum $P$ causes a
change of $\mu$. In fact, when $P$ is increased
from $P=P_{\rm{edge}}$ to larger values, the band is tilted, and the
top of the band moves upwards; the chemical potential $\mu$ should also
increase to compensate for the loss of states available, as shown 
in \reffig{qp_spectrum}(b). 
This implies $\d\mu/ \d P> 0$. On the other hand, because the system is periodic, the
existence of a branch of stationary states with $\d\mu/ \d P> 0$ at
$P=P_{\rm{edge}}$ implies the existence of another symmetric branch with
$\d\mu/ \d P< 0$ at the same point, thus suggesting the occurrence of a
swallowtail structure.

\begin{figure}[!tb,floatfix]
\centering
\resizebox{10.cm}{!} 
{\includegraphics{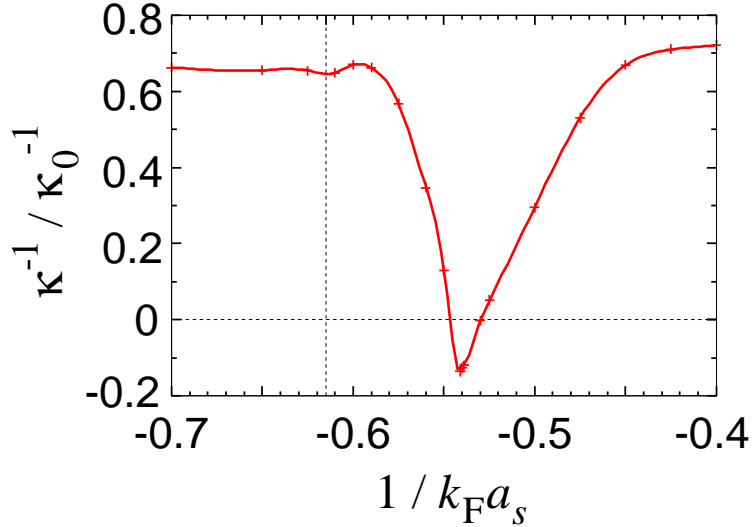}}
\caption{ Incompressibility $\kappa^{-1}$ at $P=P_{\rm{edge}}$ around the 
critical value of $1/k_{\rm F}a_s \approx -0.62$ where the swallowtail starts 
to appear. The quantity $\kappa_0^{-1}$ is the incompressibility of the 
homogeneous free Fermi gas of the same average density. In both panels, we 
have used the values $s=0.1$ and $E_{\rm F}/E_{\rm R} =2.5$.
This figure is adapted from Ref.~\onlinecite{swallowtail}.
}
\fig{inv_compress}
\end{figure}

\subsection*{5.2.\quad Incompressibility}

A direct consequence of the existence of a narrow band in the
quasiparticle spectrum near the chemical potential is a strong reduction
of the incompressibility $\kappa^{-1} = n \d \mu(n)/\d n$ close to the
critical value of $1/k_{\rm F}a_s$ in the region where swallowtails start to appear
in the BCS side (see \reffig{inv_compress}). A dip
in $\kappa^{-1}$ occurs in the situation where the top of the narrow band
is just above $\mu$ for $P=P_{\rm edge}$
($1/k_{\rm F}a_s$ is slightly above the critical value). 
An increase in the density $n$ has
little effect on $\mu$ in this case because the density of states is
large in this range of energy and the new particles can easily adjust
themselves near the top of the band by a small increase of $\mu$.
This implies that $\d \mu(n)/\d n$ is small and that the incompressibility has a
pronounced dip \cite{note_incompress}. 
It is worth noting that on the BEC side, the appearance of the swallowtail 
is not associated with any significant change in the incompressibility. 
In fact, for a Bose gas with an average density $n_{b0}$ of bosons, 
the exact solution of the GP equation gives $\kappa^{-1} = n_{b0}g$
near the critical conditions for the occurrence of swallowtails,
being a smooth and monotonic function of the interaction strength.

\subsection*{5.3.\quad Profiles of Density and Pairing Fields}

Both the pairing field and the density exhibit interesting features 
in the range of parameters where the swallowtails appear. 
This is particularly evident at the Brillouin zone boundary, $P=P_{\rm{edge}}$. 
The full profiles of $|\Delta(z)|$ and $n(z)$ 
along the lattice vector
($z$ direction) are shown in \reffig{profiles_supp}. In general, $n(z)$ and $|\Delta(z)|$ take maximum (minimum) values 
where the external potential takes its minimum (maximum) values. 
By increasing the interaction parameter $1/k_{\rm F}a_s$, we find that
the order parameter $|\Delta|$ at the maximum ($z=\pm d/2$) of the lattice
potential exhibits a transition from zero to nonzero values at the
critical value of $1/k_{\rm F}a_s$ at which the swallowtail appears.
Note that here we plot the absolute value of $\Delta$; the order
parameter $\Delta$ behaves smoothly and changes sign.

\begin{figure}[!b,floatfix]
\centering
\resizebox{8cm}{!} 
{\includegraphics{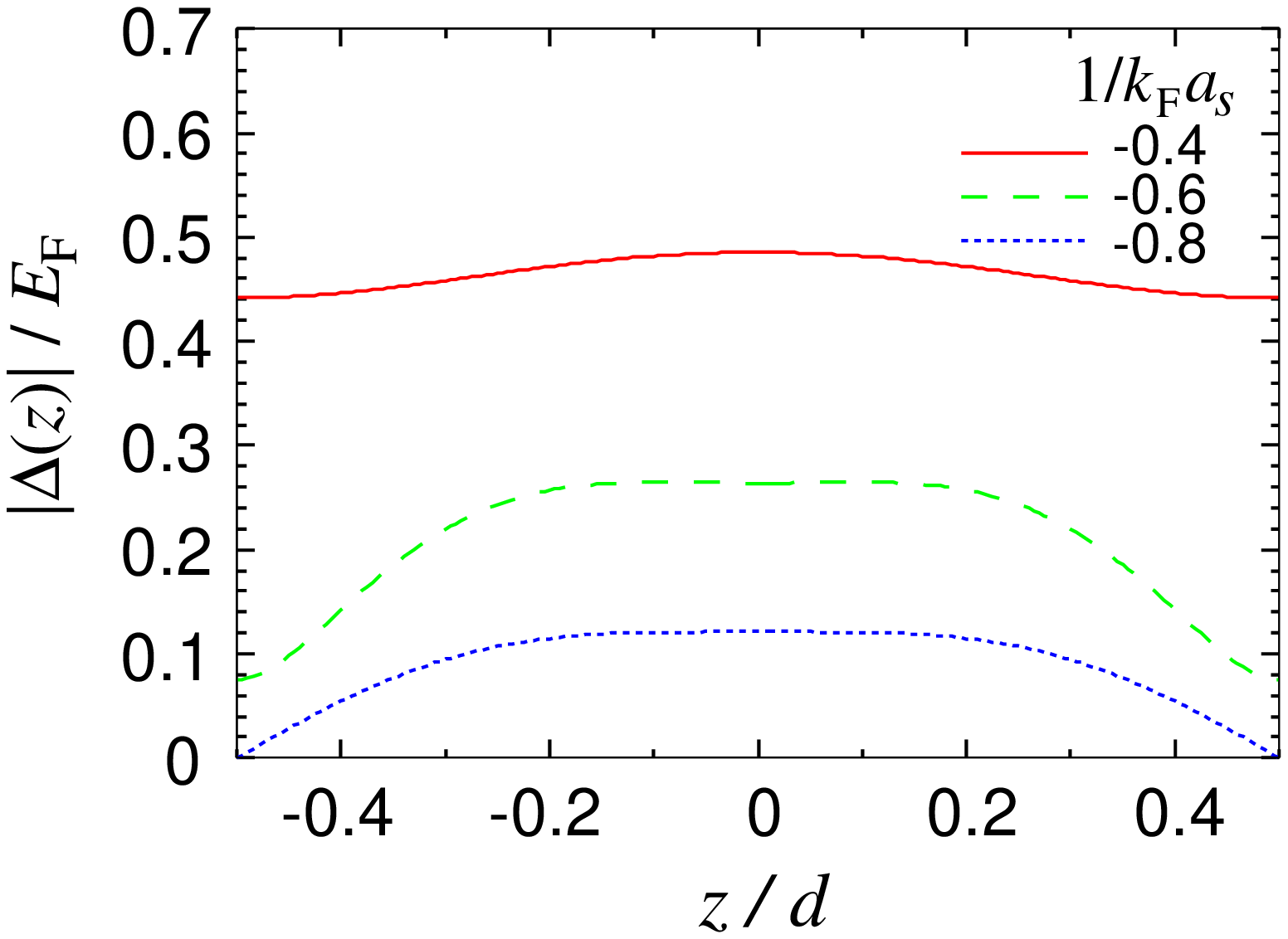}}
\resizebox{8cm}{!} 
{\includegraphics{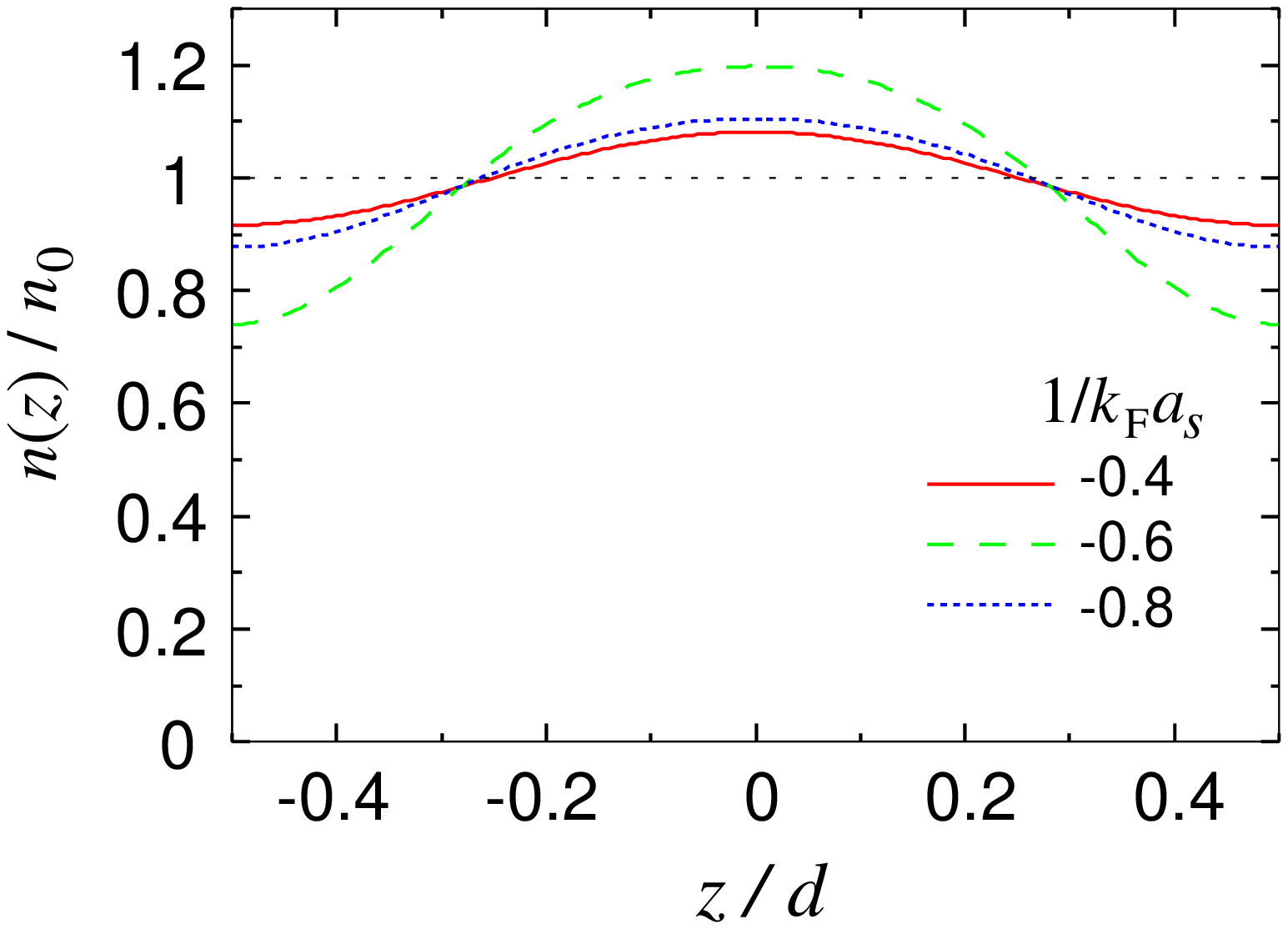}}
\caption{Profiles of the pairing field $|\Delta(z)|$ and the density $n(z)$ at
  $1/k_{\rm F}a_s= -0.8$ (blue dotted), $-0.6$ (green dashed), and $-0.4$
  (red solid) for $P=P_{\rm{edge}}$ in the case of $s=0.1$ and 
  $E_{\rm F}/E_{\rm R} =2.5$.  The swallowtail starts to appear
  at a critical value of $1/k_{\rm F}a_s \approx -0.62$.
  This figure is taken from Ref.~\onlinecite{swallowtail}.
}
\fig{profiles_supp}
\end{figure}

\begin{figure}[!tb,floatfix]
\centering
\rotatebox{270}{
\resizebox{!}{16cm} 
{\includegraphics{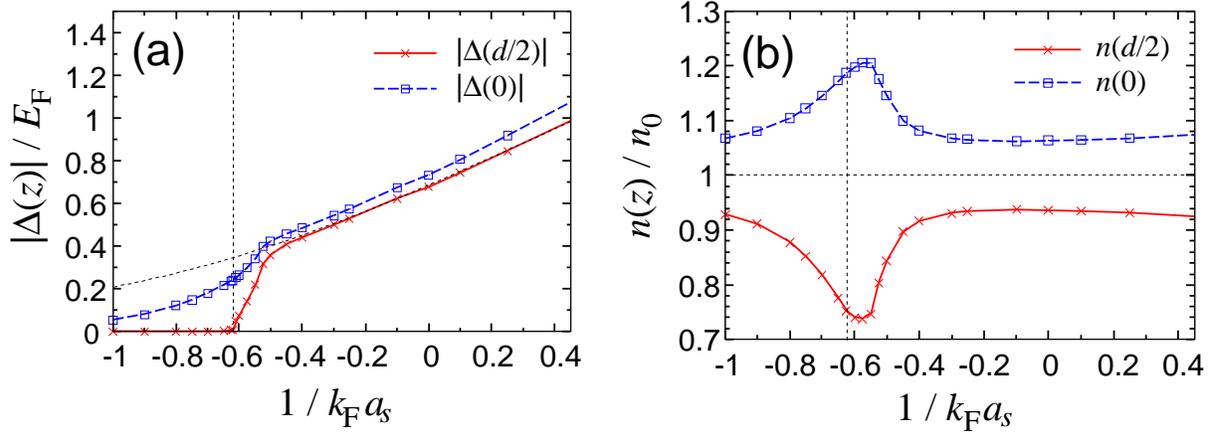}}
}
\caption{(Color online) Profiles of (a) the pairing field $|\Delta(z)|$
  and (b) the density $n(z)$ for changing $1/k_{\rm F}a_s$ for
  $P=P_{\rm{edge}}$ in the case of $s=0.1$ and $E_{\rm F}/E_{\rm R}
  =2.5$. The values of $|\Delta(z)|$ and $n(z)$ at the minimum ($z=0$,
  blue $\square$) and at the maximum ($z=\pm d/2$, red $\times$) of
  the lattice potential are shown.  The vertical dotted lines show the
  critical value of $1/k_{\rm F}a_s$ above which the swallowtail
  exists. The dotted curve in (a) shows $|\Delta|$ in a uniform system.
  This figure is taken from Ref.~\onlinecite{swallowtail}.
}
\fig{profiles}
\end{figure}

In \reffig{profiles}, we show the magnitude of the pairing field $|\Delta(z)|$ and 
the density $n(z)$ calculated at the minimum ($z=0$) 
and at the maximum ($z=\pm d/2$) of the lattice potential. 
The figure shows that $|\Delta(d/2)|$ remains zero in the BCS region 
until the swallowtail appears at $1/k_{\rm F}a_s \approx -0.62$. 
Then, it increases abruptly to values comparable to
$|\Delta(0)|$, which means that the pairing field becomes almost uniform 
at $P=P_{\rm{edge}}$ in the presence of swallowtails. As regards the density,
we find that the amplitude of the density variation, $n(0)-n(d/2)$, 
exhibits a pronounced maximum near the critical value of $1/k_{\rm F}a_s$.  

In contrast, on the BEC side,
the order parameter and the density are smooth monotonic 
functions of the interaction strength even in the region where the swallowtail 
appears. At $P=P_{\rm{edge}}$, the solution of the GP equation for bosonic
dimers gives the densities $n_b(0)= n_{b0}(1+V_0/2n_{b0} g_b)$ and 
$n_b(d/2)=n_{b0}(1-V_0/2n_{b0} g_b)$, with $V_0/2n_{b0}g_b = (3\pi/4)
(sE_{\rm R}/E_{\rm F})(1/k_{\rm F}a_s)$, where 
$g_b \equiv 4\pi \hbar^2 a_b/m_b$, and $a_b$ and $m_b$ are the scattering length
and the mass of bosonic dimers, respectively \cite{wu_st,bronski,note-nb}.  
Near the critical value of $1/k_{\rm F} a_s$, unlike the BCS side, the 
nonuniformity just decreases all the way even after the swallowtail appears. 
The local density at $z=d/2$ is zero until the swallowtail appears on the BEC
side while it is nonzero on the BCS side irrespective of the existence of the swallowtail.
The qualitative behavior of $|\Delta(z)|$ around the critical point of $1/k_{\rm F}a_s$ 
is similar to that of $n_b(z)$ 
because $n_b(\mathbf r) = (m^2 a_s/8\pi)|\Delta(\mathbf r)|^2$ \cite{bdgtogp} in the BEC limit.

\section{QUANTUM PHASES OF COLD ATOMIC GASES IN OPTICAL LATTICES \lasec{others}}

So far, we have discussed some superfluid features of cold atomic
gases in an optical lattice by mainly focusing on our works. In this
section, we discuss some important topics regarding various quantum
phases that the cold atomic gases on optical lattices can show/simulate.  
In this section, strong enough periodic potentials to
ignore the next-nearest hopping and small average numbers of particles
per lattice site will be assumed so that the system can be described by
the Hubbard lattice model.

\subsection*{6.1.\quad Quantum Phase Transition from a Superfluid to a Mott Insulator}

At a temperature of absolute zero, cold atomic gases in optical
lattices can undergo a quantum phase transition from superfluid to
Mott insulator phases as the interaction strength between atoms is
tuned from the weakly- to the strongly-interacting region \cite{Jaksch}.  The
quantum phase transition of an interacting boson gas in a periodic
lattice potential can be captured by the following Bose-Hubbard
Hamiltonian:
\beq
H = -J \sum_{\langle i,j \rangle} (\hat{b}_i^{\dagger} \hat{b}_j +\text{h.c.})+ \frac{U}{2} \sum_{i} \hat{n}_i(\hat{n}_i-1) \, , \nn
\eeq
where $\hat{b}_i$ and $\hat{b}_i^{\dagger}$ are annihilation and creation
operators of a bosonic atom on the $i$-th lattice site and $\hat{n}_i
= \hat{b}_i^{\dagger}\hat{b}_i$ is the atomic number operator on the
$i$-th site. The first term is the kinetic energy term whose strength
is characterized by the hopping matrix element $J$ between adjacent
sites $\langle i,j \rangle$, and $U(>0)$ in the second term is the
strength of a short-range repulsive interactions between bosonic
atoms.

In the limit where the kinetic energy dominates, 
the many-body ground state of $N$ atoms and $M$ lattice sites is a superposition of delocalized Bloch states with lowest energy (quasi-momentum ${\mathbf k}=0$):
\beq
\ket{ \Psi_{\text{SF}} }_{U/J=0} \propto ( \hat{b}_{ {\mathbf k}=0}^{\dagger} )^N \ket{0}  \propto \left (  \sum_{i=1}^{M} \hat{b}_i^{\dagger} \right )^N \ket{0} \, , 
\eeq
where $\ket{0}$ is the empty lattice. This state has perfect phase
correlation between atoms on different sites while the number of atoms
on each site is not fixed. This superfluid phase has gapless phonon
excitations.

In the limit where the interactions dominate (so-called, ``atomic
limit''), the fluctuations in the local atom number become
energetically unfavorable, and the ground state is made up of localized
atomic wavefunctions with a fixed number of atoms per lattice
site. The many-body ground state with a commensurate filling of $n$
atoms per lattice site is given by
\beq
\ket{ \Psi_{\text{MI}} }_{J/U=0} \propto  \prod_{i=1}^{M} (\hat{b}_i^{\dagger})^n  \ket{0}\, .
\eeq
There is no phase correlation between different sites because the
energy is independent of the phases of the wavefunctions on each
site. This Mott insulator state, unlike the superfluid state, cannot
be described by using a macroscopic wavefunction. The lowest excited state
can be obtained by moving one atom from one site to another, which
gives an energy gap of $\Delta = (U/2) [(n+1)^2+(n-1)^2-2 n^2] = U$.

As the ratio of the interaction term to the tunneling term increases
($U/J$ can be controlled by changing the depth $V_0$ of the optical
lattice even without using the Feshbach resonances), the system will
undergo a quantum phase transition from a superfluid state to a
Mott insulator state accompanying the opening of the energy gap in the
excitation spectrum.  Greiner \etal realized experimentally a
quantum phase transition from a Bose-Einstein condensate of
$^{87}\text{Rb}$ atoms with weak repulsive interactions to a Mott
insulator in a three-dimensional optical lattice potential
\cite{Greiner_SF_Mott}. Notably, they could induce reversible changes
between these two ground states by changing the strength $V_0$ of the
optical lattice.  The superfluid-to-Mott-insulator transitions were
also achieved in one- and two-dimensional cold atomic Bose gases
\cite{Mott_dim}.
The Mott insulator phases of atomic Fermi gases with {\it repulsive}
interactions on a three-dimensional optical lattice have been
realized, and the entrance into the Mott insulating states was observed
by verifying vanishing compressibility and by measuring the
suppression of doubly-occupied sites
\cite{Jordens_Mott,Schneider_Mott}.

\subsection*{6.2.\quad Quantum Phase Transition from a Paramagnet to an Antiferromagnet}

Sachdev \etal showed that the one-dimensional Mott insulator of
spinless bosons in a tilted optical lattice can be mapped onto a
quantum Ising chain \cite{Sachdev2002}. The Bose-Hubbard Hamiltonian
for a tilted optical lattice takes the form
\begin{equation}
H = - J \sum_{i}(  \hat{b}_{i} \hat{b}_{i+1}^{\dagger}+ \hat{b}_{i}^{\dagger} \hat{b}_{i+1} )+ \frac{U}{2} \sum_{i} \hat{n}_i(\hat{n}_i-1)  -  E \sum_{i} i\hat{n}_i  \ ,\nn
\end{equation}
where $E$ is the lattice potential gradient per lattice spacing.  
For a tilt near $E=U$, the energy cost of moving an atom to its neighbor
(from site $i$ to site $i+1$) is zero. If we start with a Mott
insulator with a single atom per site, an atom can resonantly tunnel
into the neighboring site to produce a dipole state (at the link)
consisting of a quasihole-quasiparticle pair on nearest neighbor
sites. Only one dipole can be created per link, and neighboring links
cannot support dipoles together. This nearest-neighbor constraint is
the source of the effective dipole-dipole interaction that results in
a density wave ordering. If a dipole creation operator
$\hat{d}_i^{\dagger} = \hat{b}_i \hat{b}_{i+1}^{\dagger}/\sqrt{2}$ is defined,
the Bose-Hubbard Hamiltonian is mapped onto the dipole Hamiltonian:
\beq
H = -\sqrt{2}J \sum_{i} (\hat{d}_i^{\dagger} + \hat{d}_i) + (U-E) \sum_{i} \hat{d}_i^{\dagger} \hat{d}_i ,\nn
\eeq
with the constraint $\hat{d}_i^{\dagger} \hat{d}_i \le 1$ and $\hat{d}_{i}^{\dagger} \hat{d}_{i}\hat{d}_{i+1}^{\dagger} \hat{d}_{i+1} = 0$.
If the dipole present/absent link is identified with a pseudospin up/down, $\hat{S}_{i}^{z} = \hat{d}_{i}^{\dagger} \hat{d}_{i} - 1/2 $,  the pseudospin-1/2 Hamiltonian takes the form of a quantum Ising chain:
\begin{eqnarray}
H &=& J_S \sum_{i} \hat{S}_{i}^{z} \hat{S}_{i+1}^{z}  -2\sqrt{2}J \sum_{i} \hat{S}_i^{x} +(J_S - D)\sum_{i} \hat{S}_i^{z}    \nn   \\
    &=& J_S \sum_{i} ( \hat{S}_{i}^{z} \hat{S}_{i+1}^{z} - h_x \hat{S}_i^{x} + h_z \hat{S}_i^{z} )\, ,  \nn
\end{eqnarray}
where $D= E - U$ and $\sum_{i} J_S(\hat{S}_{i}^{z} +
1/2)(\hat{S}_{i+1}^{z} + 1/2) $, with $J_S \to \infty$ being added to the
Hamiltonian for implementing the constraint $\hat{d}_{i}^{\dagger}
\hat{d}_{i}\hat{d}_{i+1}^{\dagger} \hat{d}_{i+1} = 0$.  The
dimensionless transverse and longitudinal fields are defined as $h_x =
2^{3/2} J/J_S$ and $h_z = 1-D/J_S$, respectively.

Thus, a quantum phase transition from a paramagnetic phase ($D\lesssim 0$)
to an antiferromagnetic phase ($D\gtrsim 0$) can be studied by changing the
lattice potential gradient $E$ between adjacent sites. With
$^{87}\text{Rb}$ atoms, a quantum simulation of antiferromagnetic spin
chains in an optical lattice was done, and a phase transition to the
antiferromagnetic phase from the paramagnetic phase was observed
\cite{Ising_chain}.

\section{SUMMARY AND OUTLOOK}
\lasec{summary}

In this article, we have focused on some superfluid properties of
cold atomic gases in an optical lattice periodic along one spatial
direction: basic macroscopic and static properties, the stability of the
superflow, and peculiar energy band structures. As a complement, some
phases other than the superfluid phase have been discussed in the last
section. Considering the rapid growth and interdisciplinary nature of
the research on cold atomic gases in optical lattices, it is
practically impossible to cover all perspectives.
We conclude by mentioning a few of them.
 
The high controllability of
the lattice potential opens many possibilities: An optical disordered
lattice can be constructed to study the problem of the Anderson
localization of matter waves and the resulting phases
\cite{disorder}. The time modulation of the optical lattice is shown
to tune the magnitude and the sign of the tunnel coupling in the Hubbard
model, which allows us to study various phases
\cite{time_modulation}. Cold atomic gases in optical lattices may
uncover many exotic phases that are still under debate or even lack
solid state analogs.  

Due to their long characteristic time scales and large characteristic length scales,
cold atomic gases are good playgrounds for the experimental observation
and control of their dynamics. Particularly, a sudden quench can be
realized experimentally. The nonequilibrium dynamics after sudden
quenches can be studied with high precision in experiments to
discriminate between candidate theories. Long-standing problems
such as thermalization, its connection with non-integrability and/or
quantum chaos are actively sought-after topics in this direction
\cite{non-equilibrium}.

The charge neutrality of atoms seems to limit the use of this
system as a quantum simulator. However, the internal states of an atom,
together with atom-laser interactions, can be exploited for the atom to
gain geometric Berry's phases, which amounts to generating artificial
gauge fields interacting with these charge-neutral particles
\cite{artificial_gauge}. In this way, interactions with
electromagnetic fields, spin-orbit coupling, and even non-Abelian
gauge fields can be emulated to open the door to the study of the physics of
the quantum Hall effects, topological superconductors/insulators, and high
energy physics by using cold atoms in the future.

Cold atomic gases provide a promising platform for controlling
dissipation and for engineering the Hamiltonian, e.g., by
controlling the coupling with a subsystem acting as a reservoir, by
using external fields that induce losses of trapped atoms, etc.  This
possibility will allow us to use the cold atomic gases as quantum
simulators of open systems.  In addition, the controlled dissipation
will offer an opportunity to study the quantum dynamics driven by
dissipation and its steady states, and to study the non-equilibrium phase transitions
among the steady states determined by the competition between the
coherent and the dissipative dynamics.  Furthermore, controlling
the dissipation will pave the way to the design of Liovillian and
to the dissipation-driven state-preparation (e.g., Ref.~\onlinecite{dissipation}).

Last, but not least, cold atomic gases in optical
lattices are also useful for the precision measurements; 
the ``optical lattice clock'' consists of millions
of atomic clocks trapped in an optical lattice and working in parallel.
The large number of simultaneously-interrogated atoms greatly improves the
stability of the clock, and state-of-the-art optical lattice clocks 
outperform the primary frequency standard of Cs clocks
(Ref.\ \onlinecite{optlatclock} and references therein).  
Bloch oscillations of cold atomic gases in optical
lattices offer a promising way of measuring forces 
at a spatial resolution of few micrometers 
(e.g., Ref.\ \onlinecite{bloch_osc}).
Precision measurement devices/techniques will enable high-precision tests of time and space variations of the fundamental constants,
the weak equivalence principle, and Newton's gravitational law at
short distances.

\begin{acknowledgments}
We acknowledge Mauro Antezza, Franco Dalfovo, Elisabetta Furlan,
Giuliano Orso, Francesco Piazza, Lev P. Pitaevskii, and Sandro Stringari
for collaborations.
This work was supported in part by the Max Planck Society, by the Korean
Ministry of Education, Science and Technology, by Gyeongsangbuk-Do,
by Pohang City [support of the Junior Research Group (JRG) at 
the Asia Pacific Center for Thoretical Physics (APCTP)], and by Basic Science
Research Program through the National Research Foundation of Korea
(NRF) funded by the Ministry of Education, Science and Technology
(No. 2012R1A1A2008028).
Calculations were performed by the RIKEN Integrated Cluster of
Clusters (RICC) system, by WIGLAF at the University of Trento, 
and by BEN at ECT*.  
\end{acknowledgments}


\begin{thebibliography}{99}
\bibitem{note_nbar} Here, the coarse-grained density $\bar{n}$ and 
the averaged quasi-momentum $\bar{P}$ are defined
as the density and the quasi-momentum averaged over the unit cell.
\bibitem{kink} We observe that, in the LDA hydrodynamic theory,
$n(z)$ and $v(z)$ exhibit a kink at the point where $v(z)=c_s(z)$,
with a finite jump in the first derivative, and that one cannot
construct a stationary solution for $v(z)>c_s(z)$.
\bibitem{footnote_parameter_value} We mainly present the result for 
$s = V_0/E_{\rm R} = 0.1$ and $E_{\rm F}/E_{\rm R} =2.5$ as an example, 
where $E_{\rm F} = \hbar^2k_{\rm F}^2/(2m)$ 
and $k_{\rm F}=(3\pi^2 n_0)^{1/3}$ are the Fermi energy and 
the momentum of a uniform free Fermi gas of density $n_0$, respectively.
These values fall in the range of parameters of feasible experiments \cite{miller}.
\bibitem{note_incompress} With the parameters used in 
\reffig{qp_spectrum}(b), the incompressibility takes negative values
in a small region around $1/k_{\rm F}a_s =-0.55$, which means that the system 
might be dynamically unstable against long-wavelength perturbations.
If appropriate parameters are chosen, this region disappears.
\bibitem{note-nb} This result is exact only when the swallowtails exist; 
otherwise, it is approximate, but qualitatively correct.
\end{thebibliography}
\end{document}